\documentclass[aps,prb]{revtex4}

\usepackage{amssymb}
\usepackage{amsmath}
\usepackage{graphicx}

\begin{document}

\title{The extended Hubbard model in the ionic limit}

\author{F. Mancini}

\affiliation{Dipartimento di Fisica ''E.R. Caianiello" - Laboratorio
Regionale SuperMat, INFM \\ Universit\`{a} degli Studi di Salerno,
I-84081 Baronissi (SA), Italy}

\begin{abstract}
In this paper, we study the Hubbard model with intersite Coulomb
interaction in the ionic limit (i.e. no kinetic energy). It is shown
that this model is isomorphic to the spin-1 Ising model in presence
of a crystal field and an external magnetic field. We show that for
such models it is possible to find, for any dimension, a finite
complete set of eigenoperators and eigenvalues of the Hamiltonian.
Then, the hierarchy of the equations of motion closes and analytical
expressions for the relevant Green's functions and correlation
functions can be obtained. These expressions are formal because
these functions depend on a finite set of unknown parameters, and
only a set of exact relations among the correlation functions can be
derived. In the one-dimensional case we show that by means of
algebraic constraints it is possible to obtain extra equations which
close the set and allow us to obtain a complete exact solution of
the model. The behavior of the relevant physical properties for the
1D system is reported.
\end{abstract}

\date{\today}
\maketitle

\section{Introduction}

In a recent paper \cite{Mancini05} we have shown that there is a
large class of fermionic systems for which it is possible to find a
complete set of eigenoperators and eigenvalues of the Hamiltonian.
Then, the hierarchy of the equations of motion closes and analytical
expressions for the Green's functions (GF) can be obtained.

In this article, we apply this formulation to the extended Hubbard
model, where a nearest-neighbor Coulomb interaction term is added to
the original Hamiltonian. This model is one of the simplest models
capable to describe charge ordering in interacting electron systems,
experimentally observed in a variety of systems. We will study the
model in the ionic limit, where the kinetic energy is neglected with
respect to the local and intersite Coulomb interactions. Among the
many analytical methods used to study the extended Hubbard model we
recall: Hartree-Fock approximation \cite{Seo98}, perturbation theory
\cite{Vandongen94}, dynamical mean field theory \cite{Pietig99},
slave boson approach \cite{Mckenzie01,Merino01}, coherent potential
approximation \cite{Hoang02}. Numerical studies by means of Quantum
Monte Carlo \cite{Hirsch84}, Lanczos technique \cite{Hellberg01} and
exact diagonalization \cite{Calandra02} have also to be recalled.

As it will be shown in Section 2, the extended Hubbard model in the
ionic limit is isomorphic to the spin-1 Ising model in presence of a
crystal field $\Delta $ and an external magnetic field $h$. The
latter model is known as the Blume-Capel (BC) model
\cite{Blume66,Capel66,Capel67,Capel67a}. With the addition of a
biquadratic interaction $K$ it is known as the Blume-Emery-Griffiths
(BEG) model \cite{Blume71} and has been largely applied to the study
of fluid mixtures and critical phenomena. The model is also related
to the three-component model \cite{Furman77}. Some exact results for
the BC and BEG models are known. In one dimension and zero magnetic
field, the spin-1 Ising model and the BEG model have been solved
exactly by means of the transfer matrix method
\cite{Suzuki67,Hintermann69}, and by means of the Bethe method
\cite{Obokata68}. Exact solutions have also been obtained for a
Bethe lattice \cite{Chakraborty86} and for the two-dimensional
honeycomb lattice \cite{Rosengren89}. The most common approach to
the BC and BEG models is based on the use of mean field
approximation
\cite{Blume71,Bernasconi71,Mukamel74,Lajzerowicz75,Sivardiere75,Sivardiere75a,Furman77,Hoston91}.
However, renormalization group studies
\cite{Riedel74,Berker76,Burkhardt76,Burkhardt77,Mahan78,Kaufman81,Yeomans81}
show some qualitative differences from the mean field results. Among
other techniques, we mention temperature expansions
\cite{Oitmaa72,Fox72,Saul74}, cluster-variation method \cite{Lock84}
and numerical simulations \cite{Tanaka85,Wang87,Wang87a,Rachadi03}.
The self-consistent Ornstein-Zernike approximation has been used to
study the phase diagram of the 3D Blume-Capel model for spin 1
\cite{Grollau01} and spin 3/2 \cite{Grollau02}. A generalization of
the BEG model by introducing a nonsymmetric exchange interaction $L$
was introduced in Ref.~\onlinecite{Mukamel74}. The one-dimensional
case for this model was studied in Ref.~\onlinecite{Krinsky75},
where exact renormalization-group recursion relations were derived,
exhibiting tricritical and critical fixed points. Also, it should be
mentioned that the general spin-1 model can be mapped onto the
spin-1/2 Ising model, under certain constrained conditions, which
determine the corresponding subspaces of interaction parameters
($J,L,K,h,\Delta $) \cite{Mi94}. The Ising model with spin 1/2, 1
and with general spin has been also studied by means of the Green's
function formalism
\cite{Doman62,Tahir63,Doman63,Callen63,Suzuki65,Oguchi66,Tyablikov67,Tahir68,Kalashnikov69,Anderson71}.
These studies do not lead to a complete solution, but to a series of
exact relations among the spin correlation functions. These
correlation identities have been used as basis for high temperature
expansions \cite{Tahir68,Taggart79}, and in combination with the
effective-field approximation \cite{Honmura78,Siqueira85}.

The outline of the paper is as follows. In Section 2, we introduce
the model for a $d$-dimensional cubic lattice. In Section 3, we show
that it is possible to find a closed set of composite operators,
which are eigenoperators of the Hamiltonian and close the algebra.
Then, as shown in Section 4, analytical expressions of the retarded
Green's function (GF) and correlation function (CF) can be obtained.
These expressions are only formal. As the composite operators do not
satisfy a canonical algebra, the GF and CF depend on a set of
internal parameters, not calculable through the dynamics, and only
exact relations among the correlation functions are obtained. In the
framework of the Green's functions formalism, extra equations must
be found by fixing the representation. According to the scheme of
the composite operator method \cite{Mancini03,Mancini03a,Mancini04}
(COM), we fix the representation by means of the algebra (algebra
constraints). By following this scheme, in Section 5 we are able to
derive for the one-dimensional case extra equations which close the
set of relations and allow us to obtain an exact solution of the 1D
extended Hubbard model in the ionic limit. This solution is also a
solution of the 1D spin-1 Ising model in presence of a crystal field
and an external magnetic field. As already mentioned, by using the
GF formalism
\cite{Doman62,Tahir63,Doman63,Callen63,Suzuki65,Oguchi66,Tyablikov67,Tahir68,Kalashnikov69,Anderson71},
other authors have derived a set of exact equations for the
correlation functions of the Ising model. All of them did not
succeed to find the extra equations necessary to close the set.
There is one exception: in Ref.~\onlinecite{Tyablikov67} the set of
equations for the 1D spin-1/2 Ising model for an infinite chain is
closed by using ergodicity conditions for the correlation functions.
However, it should be remarked that ergodicity breaks down for
finite systems and at the critical points. In Section 6, we present
some results for the particle density, specific heat and
compressibility, both in the case of attractive and repulsive
intersite Coulomb interaction. Details of the calculations are given
in the Appendices.

We would like to comment that the extended Hubbard model, although
in the limit of localized electrons, is of physical interest. The
results reported in Section 6 show some relevant features: (a) the
behaviors of the particle density and of the double occupancy show
the occurrence, at $T=0$, of phase transitions towards charge
ordered states (in particular, for $V>0$, a checkerboard order
establishes in the region $0<\mu <4V$); (b) the specific heat
presents a double peak structure; (c) a crossing point in the
specific heat curves can be observed (it is remarkable to note that
this crossing is observed only in the region where the checkerboard
order is present and the compressibility vanishes); (d) in the low
$T$ region, the thermal compressibility exhibits a double peak
structure, with peaks localized at $\mu =0$ and $\mu =4V$. All the
above mentioned results are characteristic of the Hubbard
interactions and are somehow independent of the mobility of the
electrons. Indeed, very similar results have been obtained for the
complete Hubbard model (i.e., with finite hopping) by making use of
approximations.

\section{The model}

A simple generalization of the Hubbard model is obtained by
including an intersite Coulomb interaction. The Hamiltonian of this
model is given by
\begin{equation}
H=\sum_{\mathbf{ij}}[t_{\mathbf{ij}}-\delta _{\mathbf{ij}}\mu
]c^{\dag
}(i)c(j)+U\sum_{\mathbf{i}}n_{_{\uparrow }}(i)n_{_{\downarrow }}(i)+\frac{1}{%
2}\sum_{\mathbf{i}\neq \mathbf{j}}V_{\mathbf{ij}}n(i)n(j)
\label{2.1}
\end{equation}
with the following notation. $c(i)$ and $c^{\dag }(i)$ are
annihilation and creation operators of electrons in the spinor
notation
\begin{equation}
c(i)=\left(
\begin{array}{l}
c_{\uparrow}(i) \\
c_{\downarrow}(i)
\end{array}
\right) \;\;\;\;\;\;\;\;c^{\dag }(i)=\left(
\begin{array}{ll}
c_{\uparrow }^{\dag }(i) & c_{\downarrow }^{\dag }(i)
\end{array}
\right)  \label{2.2}
\end{equation}
and satisfy canonical anti-commutation relations:
\begin{eqnarray}
\{c_{\sigma }(\mathbf{i},t),c_{\sigma ^{\prime }}^{\dag
}(\mathbf{j},t)\}
&=&\delta _{\sigma \sigma ^{\prime }}\delta _{\mathbf{ij}}  \label{2.3} \\
\{c_{\sigma }(\mathbf{i},t),c_{\sigma ^{\prime }}(\mathbf{j},t)\}
&=&\{c_{\sigma }^{\dag }(\mathbf{i},t),c_{\sigma ^{\prime }}^{\dag }(\mathbf{%
j},t)\}=0  \nonumber
\end{eqnarray}
$\mathbf{i}$ stays for the lattice vector $\mathbf{R}_{i}$ and $i=(\mathbf{i}%
,t)$. The spinor notation will be used for all fermionic operators.
$\mu $ is the chemical potential. $t_{\mathbf{ij}}$ denotes the
transfer integral and describes hopping between different sites.
$n_{\sigma }(i)=c_{\sigma }^{\dag }(i)c_{\sigma }(i)$ is the charge
density of electrons at the site $\mathbf{i}$ with spin $\sigma $.
The strength of the local Coulomb interaction is described by the
parameter $U$. $n(i)$ is the total charge density operator
\begin{equation}
n(i)=\sum_{\sigma }c_{\sigma }^{\dag }(i)c_{\sigma }(i)=c^{\dag
}(i)c(i) \label{2.4}
\end{equation}
and $V_{\mathbf{ij}}$ describes the intersite Coulomb interaction.
In this work, we restrict the analysis to the ionic limit (i.e.,
$t_{\mathbf{ij}}=0$). By considering only first-nearest neighboring
sites, $V_{\mathbf{ij}}=-2dV\alpha _{\mathbf{ij}}$ where $d$ is the
dimensionality of the system and $\alpha _{\mathbf{ij}}$ is the
projection operator. For a cubic lattice of lattice constant $a$ we
have

\begin{equation}
\alpha _{\mathbf{ij}}=\frac{1}{N}\sum_{\mathbf{k}}e^{i\mathbf{k}\cdot (%
\mathbf{R}_{i}-\mathbf{R}_{j})}\;\;\;\;\;\;\;\;\;\;\;\;\alpha (\mathbf{k})=%
\frac{1}{d}\stackrel{d}{\sum_{n=1}}\cos (k_{n}a)  \label{2.6}
\end{equation}
Then, the Hamiltonian (\ref{2.1}) takes the form:
\begin{equation}
H=\sum_{\mathbf{i}}[-\mu n(i)+UD(i)+dVn(i)n^{\alpha }(i)]
\label{2.9}
\end{equation}
where we introduced the double occupancy operator
\begin{equation}
D(i)=n_{_{\uparrow }}(i)n_{_{\downarrow
}}(i)=\frac{1}{2}n(i)[n(i)-1] \label{2.10}
\end{equation}
Hereafter, for a generic operator $\Phi (i)$ we use the following
notation
\begin{equation}
\Phi ^{\alpha }(\mathbf{i},t)=\sum_{\mathbf{j}}\alpha _{\mathbf{ij}}\Phi (%
\mathbf{j},t)  \label{2.8}
\end{equation}

Let us consider the transformation
\begin{equation}
n(i)=1+S(i)  \label{2.11}
\end{equation}
It is clear that
\begin{equation}
\begin{array}{l}
n(i)=0 \\
n(i)=1 \\
n(i)=2
\end{array}
\quad
\begin{array}{l}
{\Leftrightarrow } \\
{\Leftrightarrow } \\
{\Leftrightarrow }
\end{array}
\quad
\begin{array}{l}
S(i)=-1 \\
S(i)=0 \\
S(i)=1
\end{array}
\label{2.12}
\end{equation}
Under the transformation (\ref{2.11}) the Hamiltonian (\ref{2.9})
can be cast in the form
\begin{equation}
H=-dJ\sum\limits_{\mathbf{i}}{}S(i)S^{\alpha }(i)+\Delta \sum\limits_{%
\mathbf{i}}{}S^{2}(i)-h\sum\limits_{\mathbf{i}}{}S(i)+E_{0}
\label{2.13}
\end{equation}
where we defined
\begin{equation}
\begin{array}{l}
{{E_{0}=(-\mu +dV)N}\hfill } \\
{{h=\mu -2dV-{\frac{1}{2}}U}}
\end{array}
\qquad
\begin{array}{l}
{J=-dV} \\
{{\Delta ={\frac{1}{2}}U}}
\end{array}
\label{2.14}
\end{equation}
Hamiltonian (\ref{2.13}) is just the spin-1 Ising model with
first-nearest neighbor interactions in presence of a crystal field
$\Delta $ and an external magnetic field $h$. We have the
equivalence
\begin{equation}
H_{Ising}=H-{{E_{0}}}  \label{2.15}
\end{equation}
The relation between the partition functions is
\begin{equation}
Z_{H}=e^{-\beta {{E_{0}}}}{Z}_{Ising}  \label{2.16}
\end{equation}
then, the thermal average of any operator A assumes the same value
in both models
\begin{equation}
\left\langle A\right\rangle _{H}=\left\langle A\right\rangle
_{Ising} \label{2.17}
\end{equation}
According to this we can choose to study either one or the other
model and obtain both solutions at once. We decided to put attention
to the model Hamiltonian (\ref{2.9}). However, all the results can
be easily extended to the Ising model by means of the property
(\ref{2.17}) and of the transformation rules (\ref{2.11}) and
(\ref{2.14}). In closing this Section, we note that the
particle-hole symmetry enjoyed by the Hubbard model corresponds to
the symmetry of the Ising model under simultaneous inversion of spin
and magnetic field.

\section{Composite fields and equations of motion}

It is immediate to see that the density operator $n_{\sigma }(i)$
does not depend on time
\begin{equation}
\text{i}{\frac{\partial }{{\partial t}}}n_{\sigma }(i)=[n_{\sigma
}(i),H]=0 \label{3.1}
\end{equation}
and the standard methods based on the equations of motion are not
applicable in terms of this operator. In order to use the Green's
function formalism, let us introduce the composite field operators
\begin{equation}
\psi _{p}^{(\xi )}(i)=\xi (i)[n^{\alpha }(i)]^{p-1}\quad \quad \quad
\psi _{p}^{(\eta )}(i)=\eta (i)[n^{\alpha }(i)]^{p-1}  \label{3.2}
\end{equation}
where $\xi (i)=[1-n(i)]c(i)$ and $\eta (i)=n(i)c(i)$ are Hubbard
operators in the spinor notation [see (\ref{2.2})]. The field
operators $\psi _{p}^{(\xi )}(i)$ and $\psi _{p}^{(\eta )}(i)$
satisfy the equations of motion
\begin{equation}
\begin{array}{l}
\text{i}{\frac{\partial }{{\partial t}}}\psi _{p}^{(\xi )}(i)=-\mu
\psi
_{p}^{(\xi )}(i)+2dV\psi _{p+1}^{(\xi )}(i) \\
\text{i}{\frac{\partial }{{\partial t}}}\psi _{p}^{(\eta )}(i)=-(\mu
-U)\psi _{p}^{(\eta )}(i)+2dV\psi _{p+1}^{(\eta )}(i)
\end{array}
\label{3.3}
\end{equation}
Apparently, the equations of motion do not constitute a closed set.
By taking higher-order time derivatives we generate a hierarchy of
composite operators. However, on the basis of the anticommutation
relations (\ref{2.3} ) the following fundamental properties of the
field $[n^{\alpha }(i)]^{p}$ can be established
\begin{equation}
\lbrack n^{\alpha
}(i)]^{p}=\sum\limits_{m=1}^{4d}{}A_{m}^{(p)}[n^{\alpha }(i)]^{m}
\label{3.4}
\end{equation}
where the coefficients $A_{m}^{(p)}$ are rational numbers which
satisfy the relation
\begin{equation}
\sum\limits_{m=1}^{4d}{}A_{m}^{(p)}=1  \label{3.5}
\end{equation}
The recurrence relation (\ref{3.4}) is proved in Appendix A. We now
define the composite operators $\psi ^{(\xi )}(i)$ and $\psi ^{(\eta
)}(i)$, multiplet operators of rank $4d+1$
\begin{equation}
\psi ^{(\xi )}(i)=\left(
\begin{array}{c}
{{{\psi _{1}^{(\xi )}(i)}}} \\
{{{\psi _{2}^{(\xi )}(i)}}} \\
\vdots \\
{{{\psi _{4d+1}^{(\xi )}}}}
\end{array}
\right) =\left(
\begin{array}{c}
{\xi (i)} \\
{{{\xi (i)[n^{\alpha }(i)]}}} \\
\vdots \\
{{{\xi (i)[n^{\alpha }(i)]^{4d}}}}
\end{array}
\right)  \label{3.7}
\end{equation}
\begin{equation}
\psi ^{(\eta )}(i)=\left(
\begin{array}{c}
{{{\psi _{1}^{(\eta )}(i)}}} \\
{{{\psi _{2}^{(\eta )}(i)}}} \\
\vdots \\
{{{\psi _{4d+1}^{(\eta )}(i)}}}
\end{array}
\right) =\left(
\begin{array}{c}
{\eta (i)} \\
{{{\eta (i)[n^{\alpha }(i)]}}} \\
\vdots \\
{{{\eta (i)[n^{\alpha }(i)]^{4d}}}}
\end{array}
\right)  \label{3.8}
\end{equation}

By means of (\ref{3.3}) and of the recurrence formula (\ref{3.4}),
these fields are eigenoperators of the Hamiltonian (\ref{2.9})
\begin{equation}
\begin{array}{l}
\text{i}{\frac{\partial }{{\partial t}}}\psi ^{(\xi )}(i)=[\psi
^{(\xi
)}(i),H]=\varepsilon ^{(\xi )}\psi ^{(\xi )}(i) \\
\text{i}{\frac{\partial }{{\partial t}}}\psi ^{(\eta )}(i)=[\psi
^{(\eta )}(i),H]=\varepsilon ^{(\eta )}\psi ^{(\eta )}(i)
\end{array}
\label{3.9}
\end{equation}
where $\varepsilon ^{(\xi )}$ and $\varepsilon ^{(\eta )}$ are the
energy matrices, of rank $(4d+1)\times (4d+1)$, which can be
calculated by means of the equations of motion (\ref{3.3}) and the
recurrence rule (\ref{3.4}). The explicit expressions of the energy
matrices are given in Appendix B. The eigenvalues $E_{n}^{(\xi )}$
and $E_{n}^{(\eta )}$ of the energy matrices are given by
\begin{equation}
\begin{array}{l}
{{E_{n}^{(\xi )}=-\mu +(n-1)V}} \\
{{E_{n}^{(\eta )}=-\mu +U+(n-1)V}\hfill }
\end{array}
\quad \{n=1,2\cdots \cdots (4d+1)\}  \label{3.12}
\end{equation}

The Hamiltonian (\ref{2.9}) has been solved since we have obtained a
closed set of eigenoperators and eigenvalues. Then, we can proceed
to the calculation of observable quantities. This will be done in
the next Sections by using the formalism of Green's functions (GF).
It is worth noticing that although at the level of equations of
motion the two fields $\psi ^{(\xi )}(i)$ and $\psi ^{(\eta )}(i)$
are decoupled, they are indeed coupled by means of the
self-consistent equations necessary to determine the correlators
appearing in the normalization matrix [see Section 4].

\section{Retarded and correlation functions}

Let us consider the retarded Green's function
\begin{equation}
G^{(ab)}(i,j)=\left\langle R[\psi ^{(a)}(i)\psi ^{(b)\dag
}(j)]\right\rangle =\theta (t_{i}-t_{j})\left\langle \{\psi
^{(a)}(i),\psi ^{(b)\dag }(j)\}\right\rangle  \label{4.1}
\end{equation}
where $\left\langle \cdots \right\rangle $ denotes the
quantum-statistical average over the grand canonical ensemble and
$a,b=\xi ,\eta $. It can be
shown that $G^{(ab)}(i,j)=\delta _{ab}\delta _{\mathbf{ij}%
}G^{(a)}(t_{i}-t_{j})$. By introducing the Fourier transform
\begin{equation}
G^{(a)}(t_{i}-t_{j})={}{\frac{\text{i}}{{(2\pi
)}}}\int\limits_{-\infty }^{+\infty }{}d\omega e^{-\text{i}\omega
(t_{i}-t_{j})}G^{(a)}(\omega ) \label{4.1b}
\end{equation}
and by means of the field equations (\ref{3.9}), the retarded GF
satisfies the equation
\begin{equation}
{\lbrack \omega -\varepsilon ^{(a)}]G^{(a)}(\omega )=I^{(a)}}
\label{4.2}
\end{equation}
where $I^{(a)}$ is the normalization matrix, defined as
\begin{equation}
I^{(a)}=\left\langle \{\psi ^{(a)}(i),\psi ^{(a)}{}^{\dag
}(i)\}\right\rangle \label{4.3}
\end{equation}
Calculations of the anticommutator for a paramagnetic phase and use
of the recursion rule (\ref{3.4}) show that the normalization matrix
has the following expression
\begin{equation}
I^{(a)}=\left(
\begin{array}{ccccc}
{{{I_{1,1}^{(a)}}}} & {{{I_{1,2}^{(a)}}}} & {\cdots } &
{{{I_{1,4d}^{(a)}}}}
& {{{I_{1,4d+1}^{(a)}}}} \\
{{{I_{1,2}^{(a)}}}} & {{{I_{1,3}^{(a)}}}} & {\cdots } & {{{I_{1,4d+1}^{(a)}}}%
} & {{{I_{2,4d+1}^{(a)}}}} \\
\vdots & \vdots & \vdots & \vdots & \vdots \\
{{{I_{1,4d}^{(a)}}}} & {{{I_{1,4d+1}^{(a)}}}} & {\cdots } & {{{\
I_{4d-1,4d+1}^{(a)}}}} & {{{I_{4d,4d+1}^{(a)}}}} \\
{{{I_{1,4d+1}^{(a)}}}} & {{{I_{2,4d+1}^{(a)}}}} & {\cdots } & {{{\
I_{4d,4d+1}^{(a)}}}} & {{{I_{4d+1,4d+1}^{(a)}}}}
\end{array}
\right)  \label{4.4}
\end{equation}
where
\begin{equation}
I_{p,4d+1}^{(a)}=\sum\limits_{m=1}^{4d}{}A_{m}^{(4d+1)}I_{p-1,m+1}^{(a)}%
\quad \quad (p=2,3,\cdots ,4d+1)  \label{4.5}
\end{equation}
We see that we need to calculate only the $4d+1$ elements $%
I_{1,p}^{(a)}\;(p=1,2,\cdots ,4d+1)$. These elements have the
following expressions
\begin{equation}
\begin{array}{l}
I_{1,p}^{(\xi )}=\kappa ^{(p-1)}-\lambda ^{(p-1)} \\
I_{1,p}^{(\eta )}(\mathbf{k})=\lambda ^{(p-1)}
\end{array}
\label{4.6}
\end{equation}
with the definitions
\begin{equation}
\begin{array}{l}
{\kappa ^{(p)}=\left\langle {[n^{\alpha }(i)]^{p}}\right\rangle } \\
{\lambda ^{(p)}={\frac{1}{2}}\left\langle {n(i)[n^{\alpha }(i)]^{p}}%
\right\rangle }
\end{array}
\label{4.10}
\end{equation}

The solution of Eq. (\ref{4.2}) is
\begin{equation}
{{G^{(a)}(\omega )=\sum\limits_{n=1}^{4d+1}{}{\frac{{\sigma ^{(a,n)}}}{{%
\omega -E_{n}^{(a)}+}\text{{i}}{\delta }}}}}  \label{4.11}
\end{equation}
where the spectral functions $\sigma _{\mu \nu }^{(a,n)}$ are
calculated by means of the formula \cite{Mancini03}
\begin{equation}
\begin{array}{l}
{\sigma _{\mu \nu }^{(a,n)}=\Omega _{\mu n}^{(a)}\sum\limits_{\delta }{}}[{%
\Omega _{n\delta }^{(a)}{}]^{-1}I_{\delta \nu }^{(a)}}
\end{array}
\label{4.12}
\end{equation}
where $\Omega ^{(a)}$ is the $(4d+1)\times (4d+1)$ matrix whose
columns are the eigenvectors of the matrix $\varepsilon ^{(a)}$.
Calculations of the matrices $\Omega ^{(\xi )}$ and $\Omega ^{(\eta
)}$ are reported in Appendix B. It is worth noting that we have
$\Omega ^{(\xi )}=\Omega ^{(\eta )}$.

The spectral density matrices $\sigma _{\mu \nu }^{(a,n)}$ are
calculated in Appendix C. They satisfy the sum rule
\begin{equation}
\sum\limits_{n=1}^{4d+1}{}\sigma _{\mu \nu }^{(a,n)}=I_{\mu \nu
}^{(a)} \label{4.12a}
\end{equation}
This is a particular case of the general sum rule
\begin{equation}
\sum\limits_{n=1}^{4d+1}{}\,[E_{n}^{(a)}]^{p}\sigma
^{(a,n)}=M^{(a,p)} \label{4.12b}
\end{equation}
where $M^{(a,p)}$ are the spectral moments defined as
\begin{equation}
M^{(a,p)}=\left\langle \{\left( \text{{i}}{\partial /\partial
t}\right) ^{p}\psi ^{(a)}(i),\psi ^{(a)\dag }(i)\}\right\rangle
\label{4.12c}
\end{equation}
The fact that the sum rule (\ref{4.12b}) is satisfied at all orders
in $p$, is a consequence of the theorem proved in
Ref.~\onlinecite{Mancini98} [see also pag. 572 in
Ref.~\onlinecite{Mancini04}]. The correlation functions can be
immediately calculate from (\ref{4.11}) by means of the spectral
theorem and are given by
\begin{equation}
C^{(a)}(t_{i}-t_{j})=\left\langle \psi ^{(a)}(\mathbf{i},t_{i})\psi
^{(a)\dag }(\mathbf{i},t_{j})\right\rangle ={}{\frac{1}{{(2\pi )}}}%
\int\limits_{-\infty }^{+\infty }{}d\omega e^{-\text{i}\omega
(t_{i}-t_{j})}C^{(a)}(\omega )  \label{4.13}
\end{equation}
with
\begin{equation}
\begin{array}{l}
{C^{(a)}(\omega )=2\pi \sum\limits_{n=1}^{4d+1}{}}\frac{{\ \sigma ^{(a,n)}}}{%
1+e^{-\beta \omega }}{\delta [\omega -E_{n}^{(a)}]}
\end{array}
\label{4.14}
\end{equation}

Equations (\ref{4.11}) and (\ref{4.14}) are an exact solution of the
model Hamiltonian (\ref{2.9}). One is able to obtain an exact
solution as the composite operators $\psi _{p}^{(\xi )}(i)$ and
$\psi _{p}^{(\eta )}(i)$ constitute a closed set of eigenoperators
of the Hamiltonian. However, as stressed in
Ref.~\onlinecite{Mancini03}, the knowledge of the GF is not fully
achieved yet. The algebra of the fields $\psi ^{(\xi )}(i)$ and
$\psi
^{(\eta )}(i)$ is not canonical: as a consequence, the normalization matrix $%
I^{(a)}$ in the equation (\ref{4.2}) contains some unknown static
correlation functions, correlators, [see Eqs.
(\ref{4.6})-(\ref{4.10})], that have to be self-consistently
calculated. According to the scheme of calculations proposed by the
composite operator method \cite{Mancini03,Mancini03a,Mancini04}, one
way of calculating the unknown correlators is by specifying the
representation where the GF are realized. The knowledge of the
Hamiltonian and of the operatorial algebra is not sufficient to
completely determine the GF. The GF refer to a specific
representation (i.e., to a specific choice of the Hilbert space) and
this information must be supplied to the equations of motion that
alone are not sufficient to completely determine the GF. The
procedure is the following. We set up some requirements on the
representation and determine the correlators in order that these
conditions be satisfied. From the algebra it is possible to derive
several relations among the operators. We will call algebra
constraints (AC) all possible relations among the operators dictated
by the algebra. This set of relations valid at microscopic level
must be satisfied also at macroscopic level, when expectations
values are considered. Use of these considerations leads to some
self-consistent equations which will be used to fix the unknown
correlator appearing in the normalization matrix. An immediate set
of rules is given by the equation
\begin{equation}
\left\langle \psi ^{(a)}(i)\psi ^{(a)\dag }(i)\right\rangle ={}{\frac{1}{{%
(2\pi )}}}\int\limits_{-\infty }^{+\infty }{}d\omega C^{(a)}(\omega
) \label{4.14a}
\end{equation}
where the l.h.s. is fixed by the AC and the boundary conditions
compatible with the phase under investigation, while in the r.h.s.
the correlation function $C^{(a)}(\omega )$ is computed by means of
the equations of motion [cfr. Eq. (\ref{4.14}].

Another important set of AC can be derived
\cite{Mancini05,Mancini05a} by observing that there exist some
operators, $O$, which project out of the Hamiltonian a reduced part
\begin{equation}
OH=OH_{0}  \label{4.14b}
\end{equation}
When $H_{0}$ and $H_{I}=H-H_{0}$ commute, the quantum statistical
average of the operator $O$ over the complete Hamiltonian $H$ must
coincide with the average over the reduced Hamiltonian $H_{0}$
\begin{equation}
Tr\{Oe^{-\beta H}\}=Tr\{Oe^{-\beta H_{0}}\}  \label{4.14c}
\end{equation}

{}Another important relation is the requirement of time
translational invariance which leads to the condition that the
spectral moments, defined by Eq. (\ref{4.12c}), must satisfy the
following relation
\begin{equation}
M_{nm}^{(ab,p)}(\mathbf{k})=[M_{mn}^{(ab,p)}(\mathbf{k})]^{*}
\label{4.14d}
\end{equation}
It can be shown that if (\ref{4.14d}) is violated, then states with
a negative norm appear in the Hilbert space. Of course the above
rules are not exhaustive and more conditions might be needed.

According to the calculations given in appendix C, the GF and the
correlation functions depend on the following parameters: external
parameters $(\mu ,T,V,U)$, internal parameters $\kappa ^{(1)},\kappa
^{(2)},\cdots ,\kappa ^{(4d)}$ and $\lambda ^{(1)},\lambda
^{(2)},\cdots ,\lambda ^{(4d)}$. By means of the algebraic relations
\begin{equation}
\begin{array}{c}
{\xi _{\uparrow }\xi _{\uparrow }^{\dag }+\eta _{\uparrow }\eta
_{\uparrow
}^{\dag }=1-n_{\uparrow }} \\
{\xi _{\downarrow }\xi _{\downarrow }^{\dag }+\eta _{\downarrow
}\eta _{\downarrow }^{\dag }=1-n_{\downarrow }}
\end{array}
\label{4.29}
\end{equation}
and by making use of the AC (\ref{4.14a}), we obtain the following
$4d+1$ self-consistent equations
\begin{equation}
C_{1,k}^{(\xi )}+C_{1,k}^{(\eta )}=\kappa ^{(k-1)}-\lambda
^{(k-1)}\quad \quad (k=1,2,\cdots 4d+1)  \label{4.30}
\end{equation}
where, recalling (\ref{4.13}) and (\ref{4.14})
\begin{equation}
\begin{array}{l}
C_{1,k}^{(a)}=\left\langle \psi ^{(a)}(i)\psi ^{(a)}{}^{\dag
}(i)\right\rangle
={{\frac{1}{2}}\sum\limits_{n=1}^{4d+1}{}T_{n}^{(a)}\sigma
_{1,k}^{(a,n)}} \\
T_{n}^{(a)}=1+\tanh \left(
{{\frac{{E_{n}^{(a)}}}{{2k_{B}T}}}}\right)
\end{array}
\label{4.31}
\end{equation}
To determine the $8d$ parameters we need other $4d-1$ equations. In
order to obtain a complete solution of the model, we must calculate
these parameters. This will be done in the next Section for the
one-dimensional case.

It is worth mentioning that the formulation given in this Section
can be easily extended to multipoint correlation functions, as
$\left\langle n(i)[n^{\alpha }(i)]^{p}n(l_{1})n(l_{2})\cdots
n(l_{s})\right\rangle $. Let us define the retarded Green's function
\begin{equation}
G^{(a,\Phi )}(t-t^{\prime })=\left\langle R[\psi
^{(a)}(\mathbf{i},t)\psi ^{(a)\dag }(\mathbf{i},t^{\prime })\Phi
]\right\rangle   \label{4.32}
\end{equation}
where $\Phi =\Phi \{n(j)\}$ is any function of the $n(j)$ with $\mathbf{j}%
\neq \mathbf{i}$. All the equations derived above remain valid by
means of the substitutions
\begin{equation}
\begin{array}{l}
I^{(a)}\longrightarrow I^{(a,\Phi )}=\left\langle \{\psi
^{(a)}(i),\psi
^{(a)}{}^{\dag }(i)\}\Phi \right\rangle  \\
{\kappa ^{(p)}\longrightarrow \kappa ^{(p,\Phi )}=\left\langle
{[n^{\alpha
}(i)]^{p}}\Phi \right\rangle } \\
{\lambda ^{(p)}\longrightarrow \lambda ^{(p,\Phi )}={\frac{1}{2}}%
\left\langle {n(i)[n^{\alpha }(i)]^{p}}\Phi \right\rangle }
\end{array}
\label{4.33}
\end{equation}
For each choice of the function $\Phi \{n(j)\}$, it is necessary to
determine the new set of parameters $\kappa ^{(p,\Phi )}$ and
$\lambda ^{(p,\Phi )}$. For example see
Ref.~\onlinecite{Mancini05a}, where  the correlation function
$\left\langle n(i)[n^{\alpha }(i)]^{p}n(j)\right\rangle $ has been
calculated for the 1D spin-1/2 Ising model.

\section{ Self-consistent equations for one-dimensional systems}

Until now the analysis has been carried on in complete generality
for a cubic lattice of $d$ dimensions. We now consider
one-dimensional systems, and in particular we will study an infinite
chain in the homogeneous phase. As shown in previous Section, the
set of self-consistent equations (\ref {4.30}) are not sufficient to
determine all the 8 internal parameters. The remaining three
equations can be derived by algebraic considerations on the basis of
the requirement (\ref{4.14c}). We start from the algebraic relations
\begin{equation}
\begin{array}{l}
{{\xi ^{\dag }(i)n(i)=0}\hfill } \\
{{\xi ^{\dag }(i)D(i)=0}}
\end{array}
\label{5.1}
\end{equation}
which imply that
\begin{equation}
\xi ^{\dag }(i)H=\xi ^{\dag }(i)H_{0}  \label{5.2}
\end{equation}
where
\begin{equation}
H_{0}=H-2Vn(i)n^{\alpha }(i)  \label{5.3}
\end{equation}
By means of the fact that $H_{0}$ commutes with $H_{I}=H-H_{0}$, the
relation (\ref{5.2}) leads to
\begin{equation}
\xi ^{\dag }(i)e^{-\beta H}=\xi ^{\dag }(i)e^{-\beta H_{0}}
\label{5.4}
\end{equation}
Then, by means of the requirement (\ref{4.14c}), the correlation function $%
C_{1k}^{(\xi )}=\left\langle \xi (i)\xi ^{\dag }(i)[n^{\alpha
}(i)]^{k-1}\right\rangle $ can be expressed as
\begin{equation}
{\frac{{C_{1k}^{(\xi )}}}{{C_{11}^{(\xi )}}}}={\frac{{C_{1k}^{(\xi ,0)}}}{{%
C_{11}^{(\xi ,0)}}}}  \label{5.5}
\end{equation}
where
\begin{equation}
C_{1k}^{(\xi ,0)}=\left\langle \xi (i)\xi ^{\dag }(i)[n^{\alpha
}(i)]^{k-1}\right\rangle _{0}  \label{5.6}
\end{equation}
and $\left\langle \cdots \right\rangle _{0}$ denotes the thermal
average with respect to $H_{0}$. In order to calculate $C_{1k}^{(\xi
,0)}$, let us define the retarded GF
\begin{equation}
\begin{array}{l}
G_{1k}^{(\xi ,0)}(t-t^{\prime })=\left\langle R[\xi (\mathbf{i},t)\xi (%
\mathbf{i},t^{\prime })]{[n^{\alpha }(i)]^{k-1}}\right\rangle _{0} \\
G_{1k}^{(\eta ,0)}(t-t^{\prime })=\left\langle R[\eta (\mathbf{i},t)\eta (%
\mathbf{i},t^{\prime })]{[n^{\alpha }(i)]^{k-1}}\right\rangle _{0}
\end{array}
\label{5.7}
\end{equation}
By means of the equations of motion
\begin{equation}
\begin{array}{l}
{\lbrack \xi (i),H_{0}]=-\mu \xi (i)} \\
{\lbrack \eta (i),H_{0}]=-(\mu -U)\eta (i)}
\end{array}
\label{5.8}
\end{equation}
we have for an homogeneous phase
\begin{equation}
{G_{1k}^{(\xi ,0)}(\omega )={\frac{{2\left\langle {[n^{\alpha }(i)]^{k-1}}%
\right\rangle _{0}-\left\langle {n(i)[n^{\alpha
}(i)]^{k-1}}\right\rangle _{0}}}{{2(\omega +\mu +}\text{{i}}{\delta
)}}}}  \label{5.9a}
\end{equation}
\begin{equation}
{G_{1k}^{(\eta ,0)}(\omega )={\frac{\left\langle {n(i)[n^{\alpha }(i)]^{k-1}}%
\right\rangle {_{0}}}{{2(\omega +\mu -U+}\text{{i}}{\delta )}}}}
\label{5.9b}
\end{equation}
Recalling the relation between retarded and correlation functions we
have
\begin{equation}
{C_{1k}^{(\xi ,0)}={\frac{{2\left\langle {[n^{\alpha }(i)]^{k-1}}%
\right\rangle _{0}-\left\langle {n(i)[n^{\alpha
}(i)]^{k-1}}\right\rangle _{0}}}{{2(1+e^{\beta \mu })}}}}
\label{5.10a}
\end{equation}
\begin{equation}
{C_{1k}^{(\eta ,0)}={\frac{\left\langle {n(i)[n^{\alpha }(i)]^{k-1}}%
\right\rangle {_{0}}}{{2(1+e^{\beta (\mu -U)})}}}}  \label{5.10b}
\end{equation}
Recalling the algebraic relations
\begin{equation}
\begin{array}{l}
{\xi _{\sigma }\xi _{\sigma }^{\dag }+\eta _{\sigma }\eta _{\sigma
}^{\dag
}=1-n_{\sigma }} \\
{\eta _{\sigma }\eta _{\sigma }^{\dag }=n_{\sigma }-n_{\uparrow
}n_{\downarrow }}
\end{array}
\label{5.11}
\end{equation}
we obtain from (\ref{5.10a}) and (\ref{5.10b})
\begin{equation}
\begin{array}{l}
\left\langle {n(i)[n^{\alpha }(i)]^{k-1}}\right\rangle {_{0}=B_{1}\left%
\langle {[n^{\alpha }(i)]^{k-1}}\right\rangle _{0}} \\
\left\langle {D(i)[n^{\alpha }(i)]^{k-1}}\right\rangle {_{0}=B_{2}\left%
\langle {[n^{\alpha }(i)]^{k-1}}\right\rangle _{0}}
\end{array}
\label{5.12}
\end{equation}
where
\begin{equation}
{B_{1}=\left\langle {n(i)}\right\rangle _{0}={\frac{{2e^{\beta \mu
}(1+e^{\beta \mu }e^{-\beta U})}}{{(1+2e^{\beta \mu }+e^{2\beta \mu
}e^{-\beta U})}}}}  \label{5.13a}
\end{equation}
\begin{equation}
{B_{2}=\left\langle {D(i)}\right\rangle _{0}={\frac{{e^{\beta (2\mu -U)}}}{{%
\ (1+2e^{\beta \mu }+e^{2\beta \mu }e^{-\beta U})}}}}  \label{5.13b}
\end{equation}
By substituting (\ref{5.12}) into (\ref{5.10a}) and (\ref{5.10b})
\begin{equation}
\begin{array}{l}
{C_{1k}^{(\xi ,0)}=(1-B_{1}+B_{2})\left\langle {[n^{\alpha }(i)]^{k-1}}%
\right\rangle _{0}} \\
{C_{1k}^{(\eta ,0)}={\frac{1}{2}}(B_{1}-2B_{2})\left\langle
{[n^{\alpha }(i)]^{k-1}}\right\rangle _{0}}
\end{array}
\label{5.14}
\end{equation}
By substituting the first equation of (\ref{5.14}) into (\ref{5.5})
we obtain
\begin{equation}
C_{1k}^{(\xi )}=C_{11}^{(\xi )}\left\langle [n^{\alpha
}(i)]^{k-1}\right\rangle _{0}  \label{5.15}
\end{equation}
Now, we observe \cite{Fedro76} that $H_{0}$ describes a system where
the original lattice is divided in two disconnected sublattices (the
chains to the left and right of the site $\mathbf{i}$). Then, in
$H_{0}$ representation, the correlation functions which relates
sites belonging to different sublattices can be decoupled:
\begin{equation}
\left\langle a(j)b(m)\right\rangle _{0}=\left\langle
a(j)\right\rangle _{0}\left\langle b(m)\right\rangle _{0}
\label{5.16}
\end{equation}
for $\mathbf{i}$ and $\mathbf{j}$ belonging to different
sublattices. By using this property and the algebraic relation
(\ref{A4}) we have
\begin{equation}
\begin{array}{l}
\left\langle {\lbrack n^{\alpha }(i)]^{2}}\right\rangle {_{0}={\frac{1}{2}}%
X_{1}+X_{2}+{\frac{1}{2}}X_{1}^{2}} \\
\left\langle {\lbrack n^{\alpha }(i)]^{3}}\right\rangle {_{0}={\frac{1}{4}}%
X_{1}+{\frac{3}{2}}X_{2}+{\frac{3}{2}}X_{1}X_{2}+{\frac{3}{4}}X_{1}^{2}} \\
\left\langle {\lbrack n^{\alpha }(i)]^{4}}\right\rangle {_{0}={\frac{1}{8}}%
X_{1}+{\frac{7}{4}}X_{2}+{\frac{9}{2}}X_{1}X_{2}+{\frac{7}{8}}X_{1}^{2}+{\
\frac{3}{2}}X_{2}^{2}}
\end{array}
\label{5.17}
\end{equation}
where we defined
\begin{equation}
\begin{array}{l}
{X_{1}=\left\langle {n^{\alpha }(i)}\right\rangle _{0}} \\
{X_{2}=\left\langle {D^{\alpha }(i)}\right\rangle _{0}}
\end{array}
\label{5.18}
\end{equation}
Then, we obtain the self-consistent equations
\begin{equation}
\begin{array}{l}
C_{14}^{(\xi )}=C_{11}^{(\xi )}[{\frac{1}{4}}X_{1}+{\frac{3}{2}}X_{2}+{\frac{%
3}{2}}X_{1}X_{2}+{\frac{3}{4}}X_{1}^{2}] \\
C_{15}^{(\xi )}=C_{11}^{(\xi )}[{\frac{1}{8}}X_{1}+{\frac{7}{4}}X_{2}+{\frac{%
9}{2}}X_{1}X_{2}+{\frac{7}{8}}X_{1}^{2}+{\frac{3}{2}}X_{2}^{2}]
\end{array}
\label{5.19}
\end{equation}
which relate the correlation functions $C_{14}^{(\xi )}$,
$C_{15}^{(\xi )}$ to $C_{11}^{(\xi )}$, $C_{12}^{(\xi )}$,
$C_{13}^{(\xi )}$, when we observe that, by means of (\ref{5.15}),
the two parameters $X_{1}$ and $X_{2}$ are
expressed in terms of the correlation functions $C_{11}^{(\xi )}$, $%
C_{12}^{(\xi )}$, $C_{13}^{(\xi )}$ as
\begin{equation}
{X_{1}={\frac{{C_{12}^{(\xi )}}}{{C_{11}^{(\xi )}}}}}  \label{5.21a}
\end{equation}
\begin{equation}
{X_{2}={\frac{{C_{13}^{(\xi )}}}{{C_{11}^{(\xi )}}}}-{\frac{1}{2}}{\ \frac{{%
C_{12}^{(\xi )}}}{{C_{11}^{(\xi )}}}}-{\frac{1}{2}}{\frac{{\
C_{12}^{(\xi )}{}^{2}}}{{C_{11}^{(\xi )}{}^{2}}}}}  \label{5.21b}
\end{equation}

We need another equation. To this purpose, we start from the
algebraic relations
\begin{equation}
\begin{array}{l}
{{D^{p}(i)=D(i)}} \\
{{D(i)n^{p}(i)=2^{p}D(i)}}
\end{array}
\qquad \quad p\ge 1  \label{5.22}
\end{equation}
 From here, with some effort, we can derive the following relations
\begin{equation}
D(i)e^{-\beta
H}=D(i)\{1+\sum\limits_{p=1}^{4}{}(2f_{p}+g_{p})[n^{\alpha
}(i)]^{p}\}e^{-\beta H_{0}}  \label{5.23}
\end{equation}
\begin{equation}
D^{\alpha }(i)e^{-\beta H}=D^{\alpha
}(i)\{1+\sum\limits_{p=1}^{4}{}[f_{p}n(i)+g_{p}D(i)][n^{\alpha
}(i)]^{p}\}e^{-\beta H_{0}}  \label{5.24}
\end{equation}
where
\begin{equation}
\begin{array}{l}
{f_{1}=2A_{1}-{\frac{{13}}{3}}A_{1}^{2}-{\frac{1}{2}}A_{2}^{2}+{\frac{8}{3}}%
A_{1}A_{2}} \\
{f_{2}={\frac{{40}}{3}}A_{1}^{2}+{\frac{{11}}{6}}A_{2}^{2}-{\frac{{28}}{3}}%
A_{1}A_{2}} \\
{f_{3}=-{\frac{{32}}{3}}A_{1}^{2}-2A_{2}^{2}+{\frac{{28}}{3}}A_{1}A_{2}} \\
{f_{4}={\frac{8}{3}}A_{1}^{2}+{\frac{2}{3}}A_{2}^{2}-{\frac{8}{3}}A_{1}A_{2}}
\end{array}
\label{5.25}
\end{equation}
\begin{equation}
\begin{array}{l}
{g_{1}=-4A_{1}+2A_{2}+{\frac{{26}}{3}}A_{1}^{2}-{\frac{{10}}{3}}A_{2}^{2}-{\
\frac{1}{2}}A_{4}^{2}-{\frac{{16}}{3}}A_{1}A_{2}+{\frac{8}{3}}A_{2}A_{4}} \\
{g_{2}=-{\frac{{80}}{3}}A_{1}^{2}+{\frac{{29}}{3}}A_{2}^{2}+{\frac{{11}}{6}}%
A_{4}^{2}+{\frac{{56}}{3}}A_{1}A_{2}-{\frac{{28}}{3}}A_{2}A_{4}} \\
{g_{3}={\frac{{64}}{3}}A_{1}^{2}-{\frac{{20}}{3}}A_{2}^{2}-2A_{4}^{2}-{\frac{%
{56}}{3}}A_{1}A_{2}+{\frac{{28}}{3}}A_{2}A_{4}} \\
{g_{4}=-{\frac{{16}}{3}}A_{1}^{2}+{\frac{4}{3}}A_{2}^{2}+{\frac{2}{3}}%
A_{4}^{2}+{\frac{{16}}{3}}A_{1}A_{2}-{\frac{8}{3}}A_{2}A_{4}}
\end{array}
\label{5.26}
\end{equation}
and
\begin{equation}
A_{p}=e^{-p\beta V}-1  \label{5.27}
\end{equation}
By taking the expectation value and by using the relations
(\ref{5.12}) and ( \ref{5.15}), we obtain from (\ref{5.23}) and
(\ref{5.24})
\begin{equation}
\left\langle D(i)\right\rangle ={\frac{{B_{2}}}{{(1-B_{1}+B_{2})}}}
\{C_{1,1}^{(\xi)}+\sum\limits_{p=1}^{4}{}(2f_{p}+g_{p})C_{1,p+1}^{(\xi)}\}
\label{5.28}
\end{equation}
\begin{equation}
\left\langle D^{\alpha }(i)\right\rangle ={\frac{{C_{11}^{(\xi
)}}}{{\ (1-B_{1}+B_{2})}}}[\left\langle D^{\alpha }(i)\right\rangle
_{0}+\sum\limits_{p=1}^{4}{}(B_{1}f_{p}+B_{2}g_{p})\left\langle
D^{\alpha }(i)[n^{\alpha }(i)]^{p}\right\rangle _{0}]  \label{5.29}
\end{equation}
The translational invariance requires that $\left\langle D^{\alpha
}(i)\right\rangle =\left\langle D(i)\right\rangle $. Then, from
(\ref{5.28}) and (\ref{5.29}) we obtain the equation
\begin{equation}
B_{2}+B_{2}\sum\limits_{p=1}^{4}{}(2f_{p}+g_{p}){\frac{{C_{1,p+1}^{(\xi
)}}}{{C_{11}^{(\xi)}}}}=X_{2}+\sum%
\limits_{p=1}^{4}{}(B_{1}f_{p}+B_{2}g_{p})<D^{\alpha }(i)[n^{\alpha
}(i)]^{p}>_{0}  \label{5.30}
\end{equation}
To calculate the correlation functions $\left\langle D^{\alpha
}(i)[n^{\alpha }(i)]^{p}\right\rangle _{0}$, we observe the
following algebraic relation which can be derived by means of
(\ref{A4}) and (\ref {5.22})
\begin{equation}
\begin{array}{l}
{D^{\alpha }(i)n^{\alpha }(i)=D^{\alpha }(i)+{\frac{1}{6}}n^{\alpha
}(i)-{\
\frac{1}{2}}[n^{\alpha }(i)]^{2}+{\frac{1}{3}}[n^{\alpha }(i)]^{3}} \\
{D^{\alpha }(i)[n^{\alpha }(i)]^{2}=D^{\alpha
}(i)+{\frac{1}{6}}n^{\alpha }(i)-{\frac{1}{3}}[n^{\alpha
}(i)]^{2}-{\frac{1}{6}}[n^{\alpha }(i)]^{3}+{\
\frac{1}{3}}[n^{\alpha }(i)]^{4}} \\
{D^{\alpha }(i)[n^{\alpha }(i)]^{3}=D^{\alpha
}(i)-{\frac{1}{3}}n^{\alpha }(i)+{\frac{7}{4}}[n^{\alpha
}(i)]^{2}-{\frac{{35}}{{12}}}[n^{\alpha
}(i)]^{3}+{\frac{3}{2}}[n^{\alpha }(i)]^{4}} \\
{D^{\alpha }(i)[n^{\alpha }(i)]^{4}=D^{\alpha }(i)-{\frac{{25}}{{12}}}%
n^{\alpha }(i)+{\frac{{205}}{{24}}}[n^{\alpha }(i)]^{2}-{\frac{{265}}{{24}}}%
[n^{\alpha }(i)]^{3}+{\frac{{55}}{{12}}}[n^{\alpha }(i)]^{4}}
\end{array}
\label{5.31}
\end{equation}
By taking the expectation value of (\ref{5.31}) with respect to
$H_{0}$ and by using the property (\ref{5.16}), we can express the
correlation functions $\left\langle D^{\alpha }(i)[n^{\alpha
}(i)]^{p}\right\rangle _{0}$ as
\begin{equation}
\begin{array}{l}
\left\langle {D^{\alpha }(i)n^{\alpha }(i)}\right\rangle {_{0}=X_{2}+{\frac{1%
}{2}}X_{1}X_{2}} \\
\left\langle {D^{\alpha }(i)[n^{\alpha }(i)]^{2}}\right\rangle {_{0}=X_{2}+{%
\ \frac{5}{4}}X_{1}X_{2}+{\frac{1}{2}}X_{2}^{2}} \\
\left\langle {D^{\alpha }(i)[n^{\alpha }(i)]^{3}}\right\rangle {_{0}=X_{2}+{%
\ \frac{{19}}{8}}X_{1}X_{2}+{\frac{9}{4}}X_{2}^{2}} \\
\left\langle {D^{\alpha }(i)[n^{\alpha }(i)]^{4}}\right\rangle {_{0}=X_{2}+{%
\ \frac{{65}}{{16}}}X_{1}X_{2}+{\frac{{55}}{8}}X_{2}^{2}}
\end{array}
\label{5.32}
\end{equation}
Then, we can we can put equation (\ref{5.30}) under the form
\begin{equation}
b_{0}+b_{1}X_{1}+b_{2}X_{2}+b_{3}X_{1}X_{2}+b_{4}X_{1}^{2}+b_{5}X_{2}^{2}=0
\label{5.33}
\end{equation}
where
\begin{equation}
\begin{array}{l}
{b_{0}=B_{2}} \\
{b_{1}=r_{1}-s_{1}+{\frac{1}{2}}(r_{2}-s_{2})+{\frac{1}{4}}(r_{3}-s_{3})+{\
\frac{1}{8}}(r_{4}-s_{4})} \\
{b_{2}=-1+r_{2}+{\frac{3}{2}}r_{3}+{\frac{7}{4}}r_{4}-s_{1}-2s_{2}-{\frac{5}{%
2}}s_{3}-{\frac{{11}}{4}}s_{4}} \\
{b_{3}={\frac{3}{2}}r_{3}+{\frac{9}{2}}r_{4}-{\frac{1}{2}}s_{1}-{\frac{5}{4}}%
s_{2}-{\frac{{31}}{8}}s_{3}-{\frac{{137}}{{16}}}s_{4}} \\
{b_{4}={\frac{1}{2}}(r_{2}-s_{2})+{\frac{3}{4}}(r_{3}-s_{3})+{\frac{7}{8}}%
(r_{4}-s_{4})} \\
{b_{5}={\frac{3}{2}}r_{4}-{\frac{1}{2}}s_{2}-{\frac{9}{4}}s_{3}-{\frac{{67}}{%
8}}s_{4}}
\end{array}
\label{5.34}
\end{equation}
\begin{equation}
\begin{array}{l}
{r_{p}=(B_{1}+2B_{2})f_{p}+2B_{2}g_{p}} \\
{s_{p}=B_{1}f_{p}+B_{2}g_{p}}
\end{array}
\label{5.35}
\end{equation}
Recalling [see (\ref{5.21a}) and (\ref{5.21b})] that the parameters
${X_{1}}$
and ${X_{2}}$ are expressed in terms of the correlation functions $%
C_{11}^{(\xi )}$, $C_{12}^{(\xi )}$, $C_{13}^{(\xi )}$, Eq.
(\ref{5.33}) gives the needed third equation

Summarizing, we have 8 self-consistent equations (\ref{4.30}),
(\ref{5.19}) and (\ref{5.33}) which will determine the 8 internal
parameters $\kappa ^{(1)},\kappa ^{(2)},\kappa ^{(3)},\kappa
^{(4)},\lambda ^{(1)},\lambda
^{(2)},\lambda ^{(3)},\lambda ^{(4)}$ in terms of the external parameters $%
\mu ,T,U$ and $V$. The set of 5 equations in (\ref{4.30}) is a
system of linear equations. This system can be analytically solved
with respect to 5 parameters and we are left with three parameters,
which are determined by the non-linear equations (\ref{5.19}) and
(\ref{5.33}). Once these parameters are known, we can calculate the
correlation functions and all the properties of the system.\

\section{Results for the one-dimensional case}

\begin{figure}[tb]
\includegraphics[width=8cm]{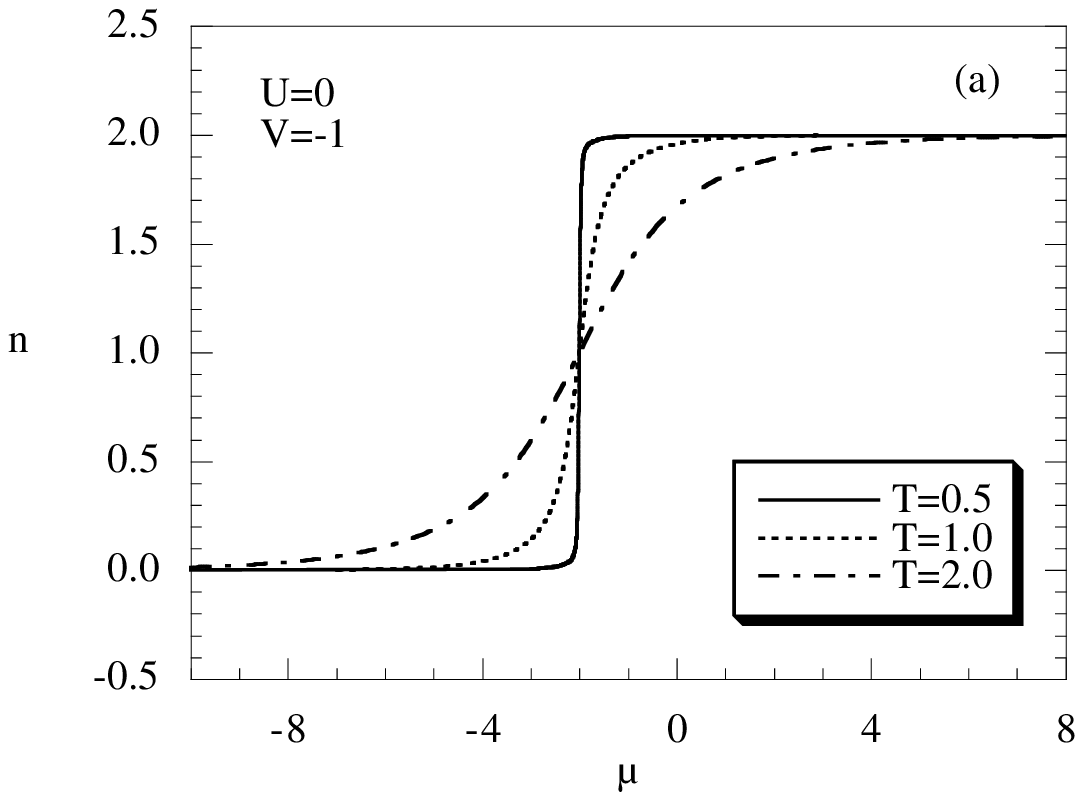}
\includegraphics[width=8cm]{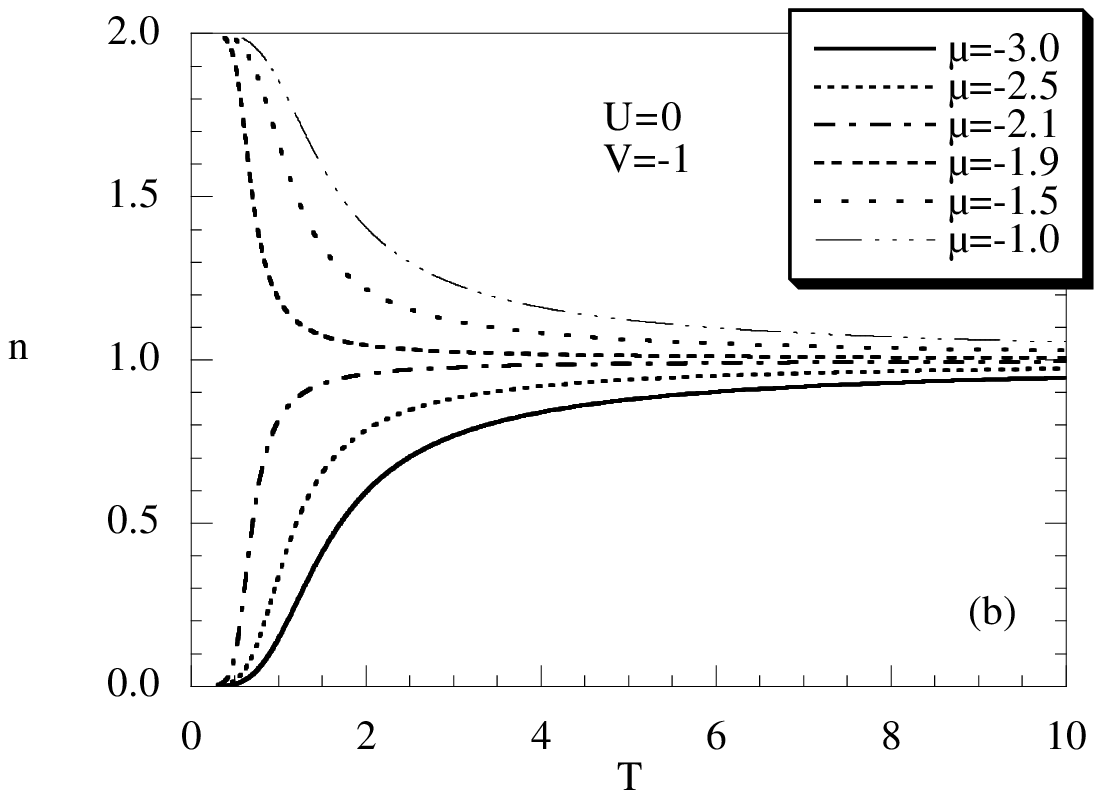}
\caption{The particle density $n$ is plotted as a function of: (a)
the chemical potential at various temperatures for $V=-1$; (b) the
temperature at various values of chemical potential for $V=-1$.}
\label{Fig1}
\end{figure}

We now present some results for the case $U=0$. This situation
corresponds to the ionic Hubbard model without local Coulomb
interaction, and to the pure spin-1 Ising model, without the crystal
field. In one dimension and zero magnetic field, the spin-1 Ising
model and the BEG model have been solved exactly by means of the
transfer matrix method
(Refs.~\onlinecite{Suzuki67,Hintermann69,Krinsky75}). We recall that
the case of zero magnetic field corresponds to the case of half
filling in the Hubbard model. The presence of magnetic field has
been treated in Ref.~\onlinecite{Krinsky75}, only for a
ferromagnetic coupling and studied only in connection to the
existence of critical and tricritical fixed points.  The general
case of $U$ different from zero will be considered elsewhere. We
study the behavior of the system
as a function of the parameters $\mu $ and $T$. We take $\left| V\right| =1$%
: all energies are measured in units of $\left| V\right| $.

At first, we consider the case of an attractive intersite Coulomb
potential (i.e., $V<0$). This situations corresponds to $J$ positive
(i.e.,
ferromagnetic coupling). In Fig.~\ref{Fig1} (a), we show the particle density $%
n=<n(i)>=\kappa ^{(1)}$ as a function of the chemical potential $\mu
$. In
terms of the Ising model, this figure should be read as the magnetization $%
\left\langle S(i)\right\rangle $ versus the magnetic field $h$. By
increasing $\mu $, the particle density increases and varies between
$0$ and $2$. At $\mu =2V$ we have $n=1$, in agreement with the
particle-hole symmetry. By decreasing the temperature, at $\mu =2V$
the system tends to become unstable against a charge ordered state
(ferromagnetic order in the Ising model): the particle density jumps
from $0$ to $2$. This is also seen in Fig.~\ref{Fig1} (b), where the
particle density is plotted versus the temperature for
various values of the chemical potential. For $\mu <2V$ we have $%
\lim_{T\rightarrow 0}n=0$, while for $\mu >2V$ we have
$\lim_{T\rightarrow 0}n=2$. At zero temperature there is a phase
transition at $\mu =2V$ from a state with no particle to a fully
occupied state where the charge assumes the maximum value.

\begin{figure}[tb]
\includegraphics[width=8cm]{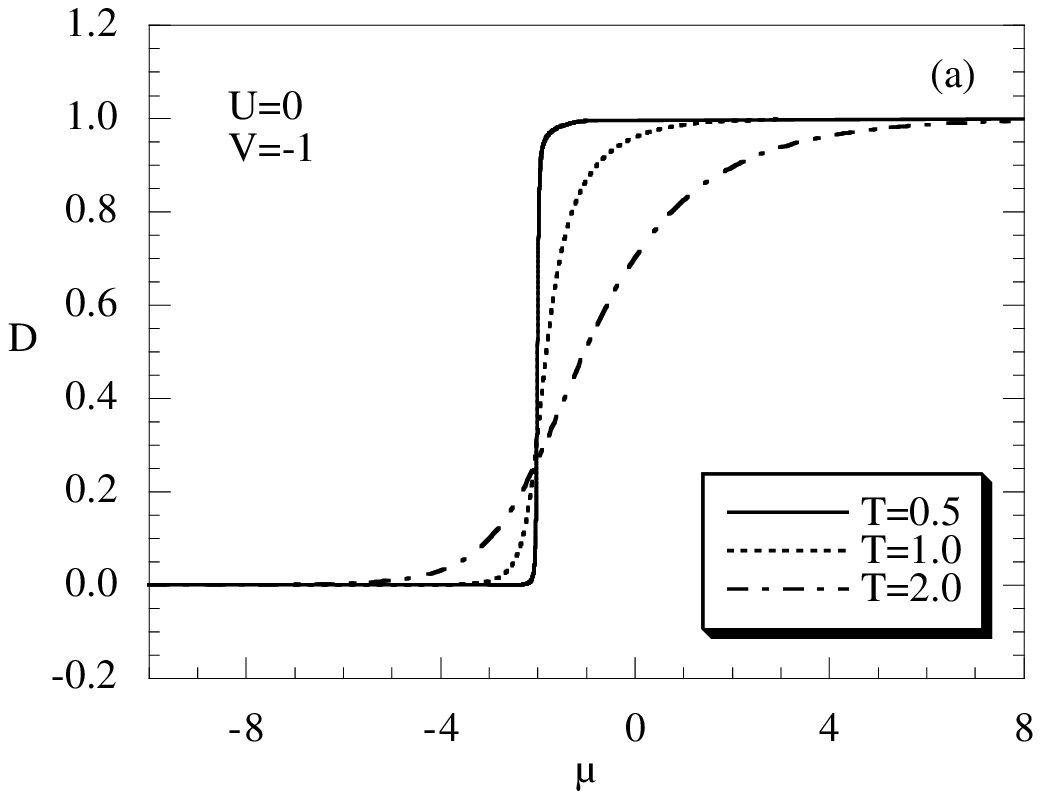}
\includegraphics[width=8cm]{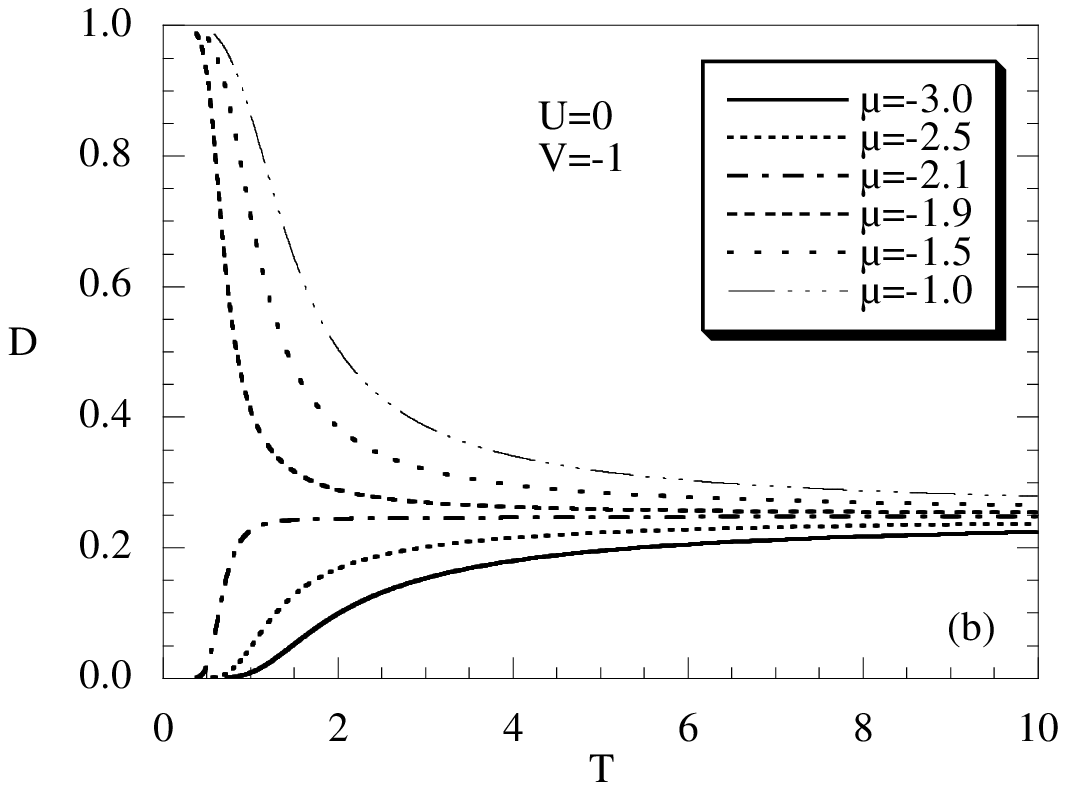}
\caption{The double occupancy $D$ is plotted as function of: (a) the
chemical potential at various temperatures for $V=-1$; (b) the
temperature at various values of chemical potential for $V=-1$.}
\label{Fig2}
\end{figure}
%

The double occupancy $D$ can be computed by means of the expression
\begin{equation}
D=\frac{n}{2}-C_{11}^{\eta \eta }  \label{6.1}
\end{equation}
The behavior of $D$ is shown in Figs.~\ref{Fig2} (a) and \ref{Fig2}
(b), where $D$ is given as a function of the chemical potential and
temperature, respectively. By increasing $\mu $, the double
occupancy increases and varies between 0 and 1. For $\mu <2V$ we
have $\lim_{T\rightarrow 0}n=0$, while for $\mu >2V$ we have
$\lim_{T\rightarrow 0}D=1$. At zero temperature there is a phase
transition at $\mu =2V$ from a state where all the sites are empty
to a state where all the sites are doubly occupied. The behavior of
the parameters $\kappa ^{(p)}$ and $\lambda ^{(p)}$ as functions of
$\mu $ is similar to that exhibited by $n$; for $T=0$ these
parameters at $\mu =2V$ jump from $0$ to their ergodic value.

\begin{figure}[tb]
\includegraphics[width=8cm]{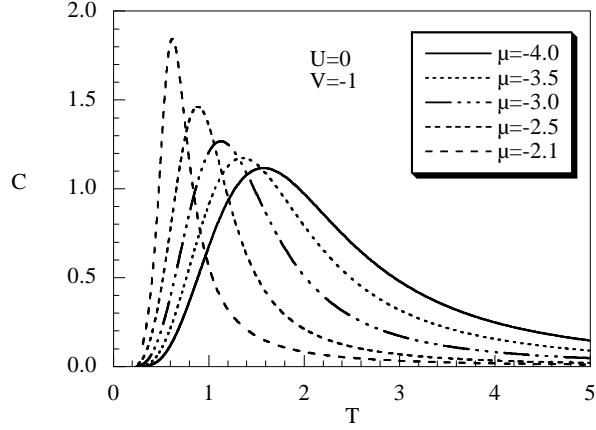}
\caption{The specific heat $C$ is plotted as a function of the
temperature at various values of chemical potential for $V=-1$.}
\label{Fig3}
\end{figure}

The specific heat is given by
\begin{equation}
C=\frac{dE}{dT}  \label{6.2}
\end{equation}
where the internal energy $E$ can be calculated by means of the
expression
\begin{equation}
E=-\mu n+UD+2V\lambda ^{(1)}  \label{6.3}
\end{equation}
The specific heat satisfies the property $C(\mu )=C(2V-\mu )$.
Therefore, we can limit the analysis to the region $2V<\mu <\infty $
(or $-\infty <\mu <2V$ ). As shown in Fig.~\ref{Fig3}, the specific
heat increases by increasing $T$ up to a
certain temperature, then decreases and goes to zero in the limit $%
T\rightarrow \infty $. Near the transition point $\mu =2V$, the peak
is
sharper and is situated in the low-temperature region. By moving away from $%
\mu =2V$, the peak becomes broader and moves to high temperatures.

The thermal compressibility $\kappa ^{T}$ is given by
\begin{equation}
\kappa ^{T}=\frac{1}{n^{2}}\frac{dn}{d\mu }  \label{6.4}
\end{equation}

\begin{figure}[tb]
\includegraphics[width=8cm]{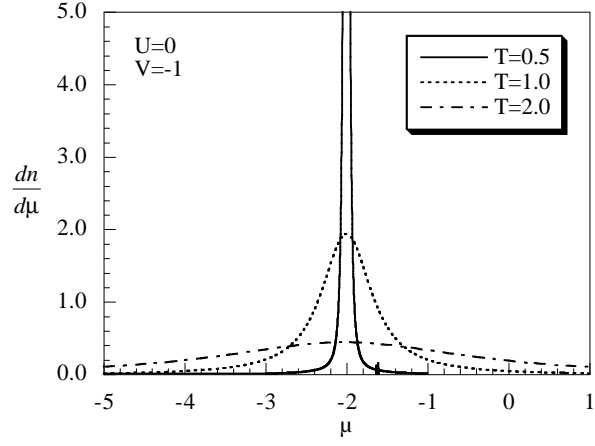}
\caption{The derivative of the particle density with respect to the
chemical potential $dn/d\mu$ is plotted as a function of the
chemical potential at various temperatures for $V=-1$.}
\label{Fig4}
\end{figure}


In Fig.~\ref{Fig4} $\frac{dn}{d\mu }$ is plotted versus the chemical
potential for
various values of temperature. A peak is observed at the transition point $%
\mu =2V$.\ By decreasing $T$, the height of the peak increases and
the compressibility tends to diverge in the limit $T\rightarrow 0$.
As a function of the temperature ($\kappa ^{T}$ $)_{\mu =2V}$
exponentially diverges at low temperatures and decreases as
$\frac{1}{T}$ in the limit of large T. In terms of the Ising model,
Fig.~\ref{Fig4} should be read as the spin susceptibility versus the
magnetic field.

\begin{figure}[tb]
\includegraphics[width=8cm]{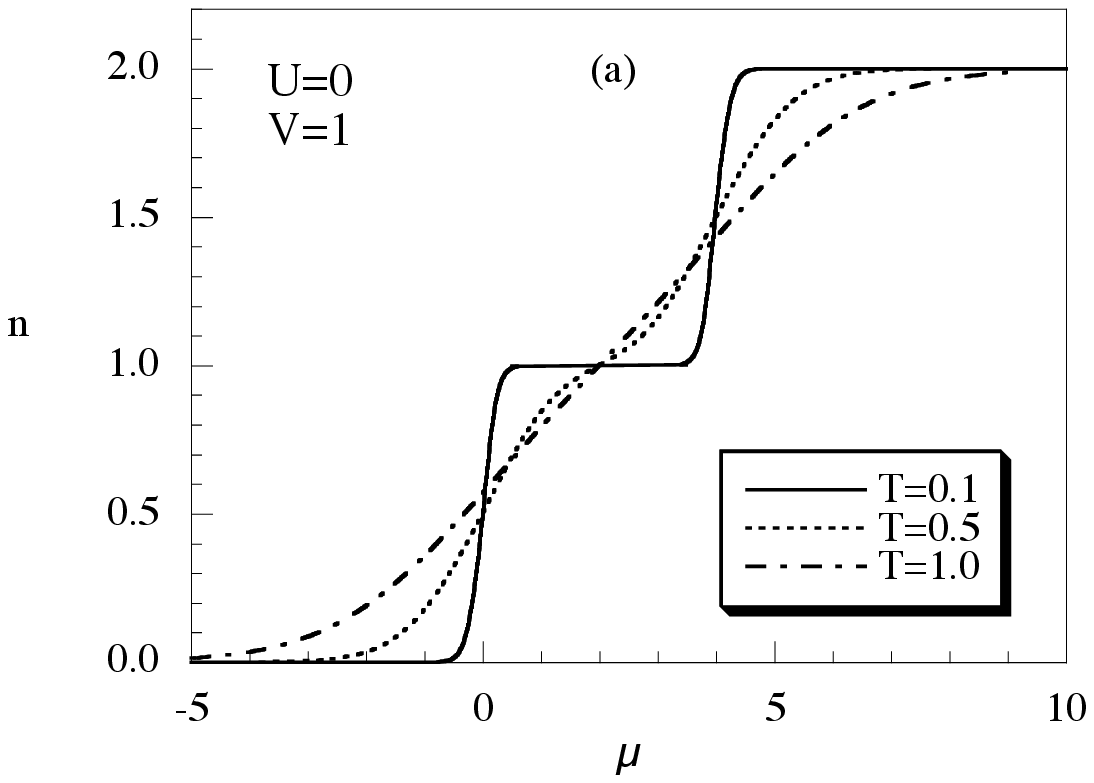}
\includegraphics[width=8cm]{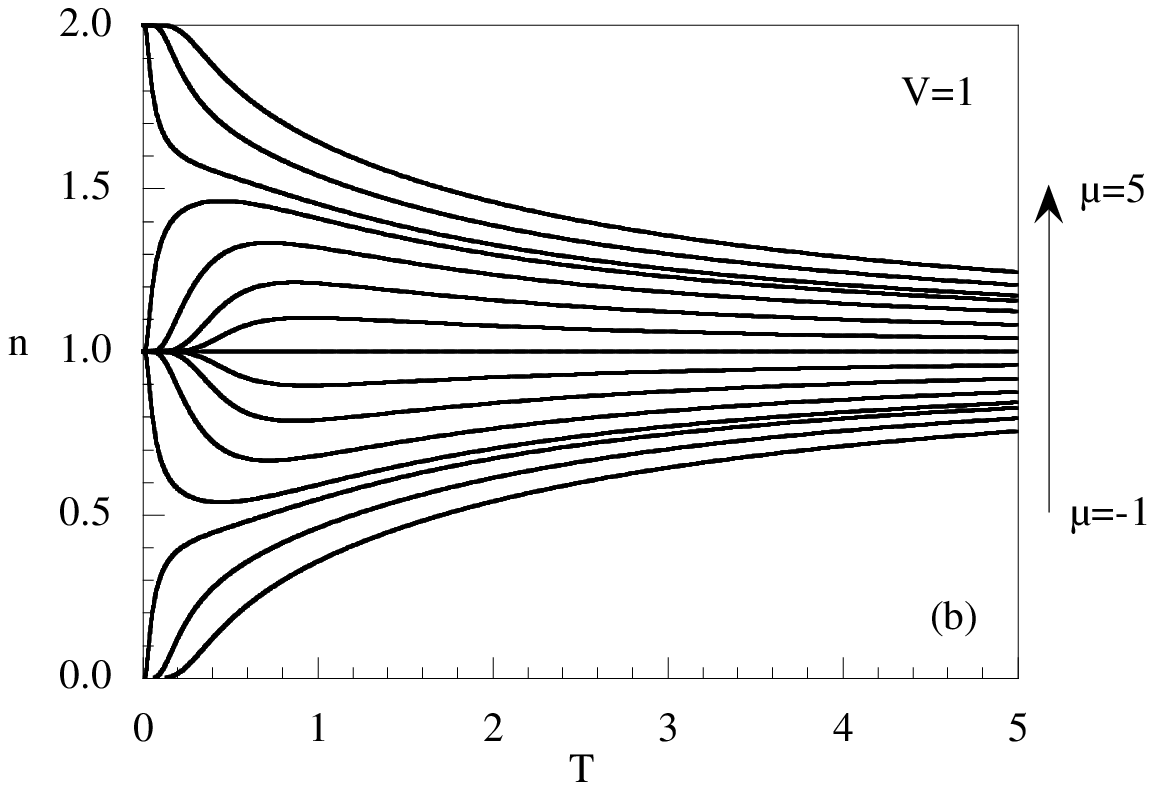}
\caption{The particle density $n$ is plotted as a function of: (a)
the chemical potential at various temperatures for $V=1$; (b) the
temperature at various values of the chemical potential for $V=1$.}
\label{Fig5}
\end{figure}

\begin{figure}[tb]
\includegraphics[width=8cm]{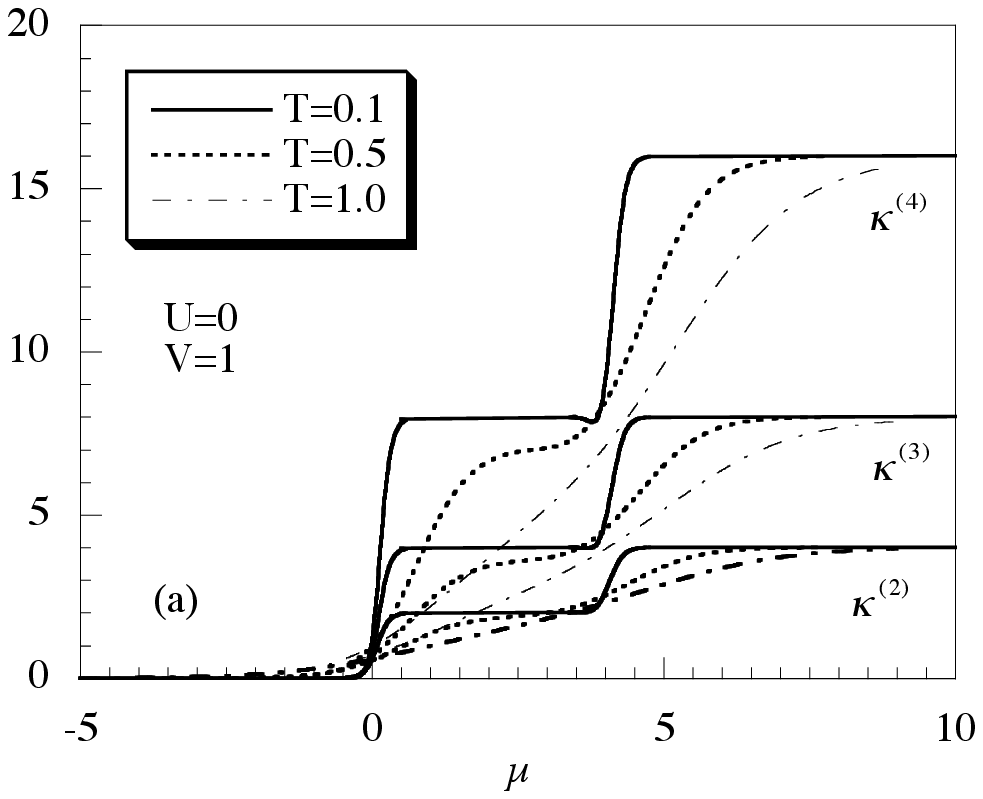}
\includegraphics[width=8cm]{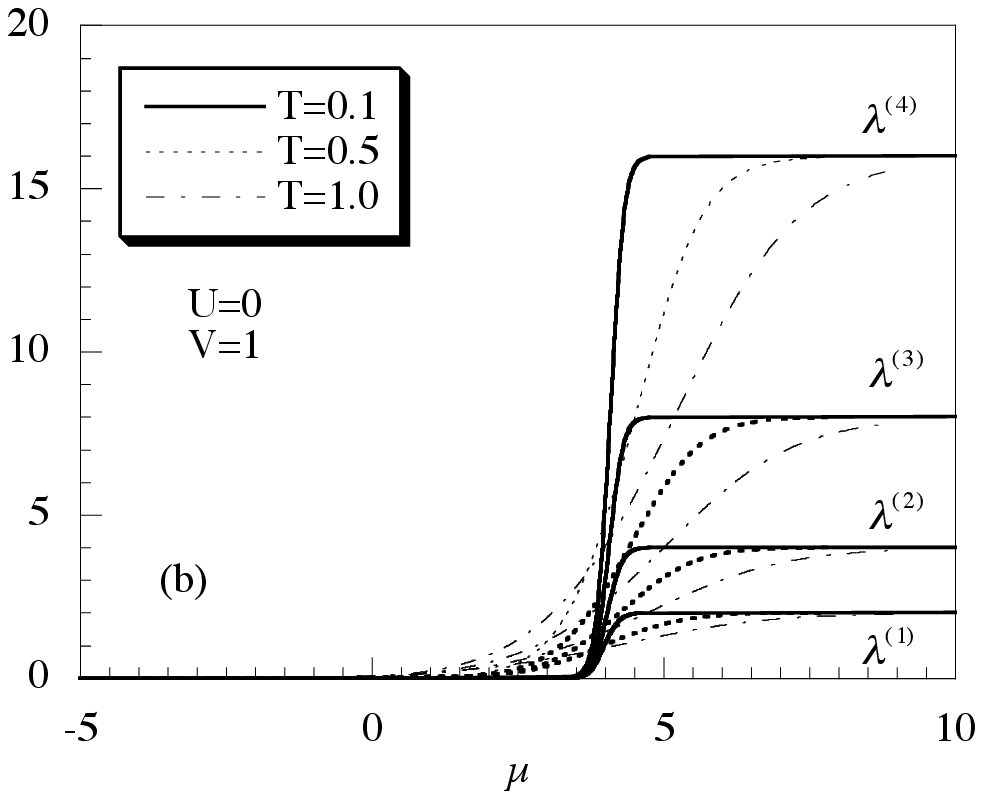}
\caption{The internal parameters (a) $\kappa^{(2)}$, $\kappa^{(3)}$, $%
\kappa^{(4)}$ and (b) $\lambda^{(1)}$, $\lambda^{(2)}$, $\lambda^{(3)}$, $%
\lambda^{(4)}$ are plotted as functions of the chemical potential at
various temperatures for $V=1$.} \label{Fig6}
\end{figure}

Next, we consider the case of a repulsive intersite Coulomb
interaction (i.e., $V>0$). This case corresponds to $J$ negative
(i.e., antiferromagnetic coupling). In Fig.~\ref{Fig5} (a) we show
the particle density $n$ as a function of the chemical potential
$\mu $. By increasing $\mu $, the particle density increases from
zero, reaches the value $1$ at $\mu =2V$, and tends to $2$ for
larger values of the chemical potential. When the temperature
decreases some instabilities of the homogeneous phase appear. In the
limit $T\rightarrow 0$, two singularities manifest: one at $\mu =0$,
where $n$ jumps from $0$ to $1$, the other at $\mu =4V$, where $n$
jumps from $1$ to $2$. In the region $0<\mu <4V$, $n$ exhibits a
plateau centered at $\mu =2V$. This behavior is also seen in
Fig.~\ref{Fig5} (b), where the particle density $n$ is given as a
function of the temperature. At $T=0$ we have two phase transitions.
At $\mu =0$ the system passes from a state with no charge
to a state where the charge is distributed in a checkerboard structure. At $%
\mu =4V$ there is a second phase transition where the system passes
from the checkerboard structure to a state where the charge is
uniformly distributed. The checkerboard structure is clearly seen
from the behavior of the parameters $\kappa ^{(p)}$ and $\lambda
^{(p)}$, as shown in Figs.~\ref{Fig6}. While the parameters $\kappa
^{(p)}$ have the same behavior as $n$, with two singularities at
$\mu =0$ and at $\mu =4V$, the parameters $\lambda ^{(p)}$ exhibit
only one singularity at $\mu =4V$. The reason of this difference is
related to the fact that $\kappa ^{(p)}$ are correlation functions
between the site $\mathbf{i}$ and second-nearest neighboring sites,
while $\lambda ^{(p)}$ mainly relate the site $\mathbf{i}$ to
first-nearest neighboring sites.

\begin{figure}[tb]
\includegraphics[width=8cm]{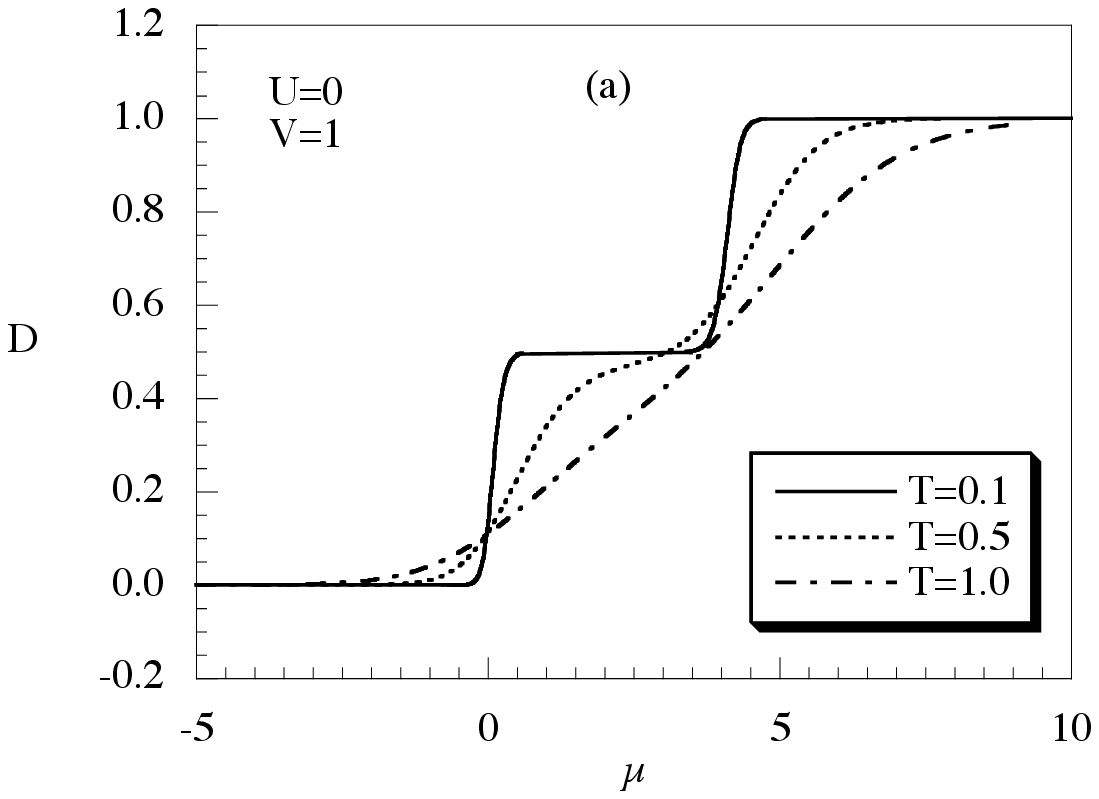}
\includegraphics[width=8cm]{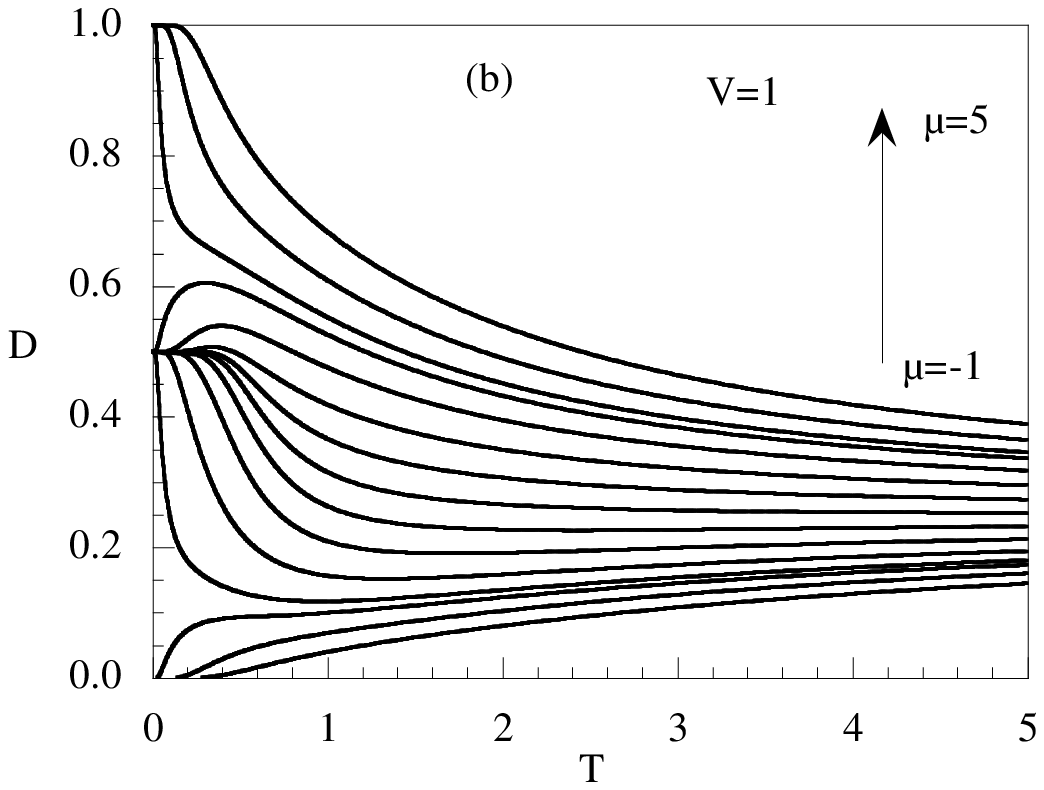}
\caption{The double occupancy $D$ is plotted as function of: (a) the
chemical potential at various temperatures for $V=1$; (b) the
temperature at various values of chemical potential for $V=1$.}
\label{Fig7}
\end{figure}

The double occupancy as a function of the chemical potential is
shown in Fig.~\ref{Fig7} (a); by increasing $\mu $, $D$ increases
from zero and tends to $1$ for large values of the chemical
potential. At $T=0$, as also seen in Fig.~\ref{Fig7} (b), $ D $ has
a discontinuity at $\mu =0$, where jumps from zero to $1/2$, and
another discontinuity at $\mu =4V$ where jumps from $1/2$ to $1$. In
the region $0<\mu <4V$ D has the constant value of $1/2$. Again,
this show the transition to a checkerboard structure of the charge.

\begin{figure}[tb]
\includegraphics[width=5.3cm]{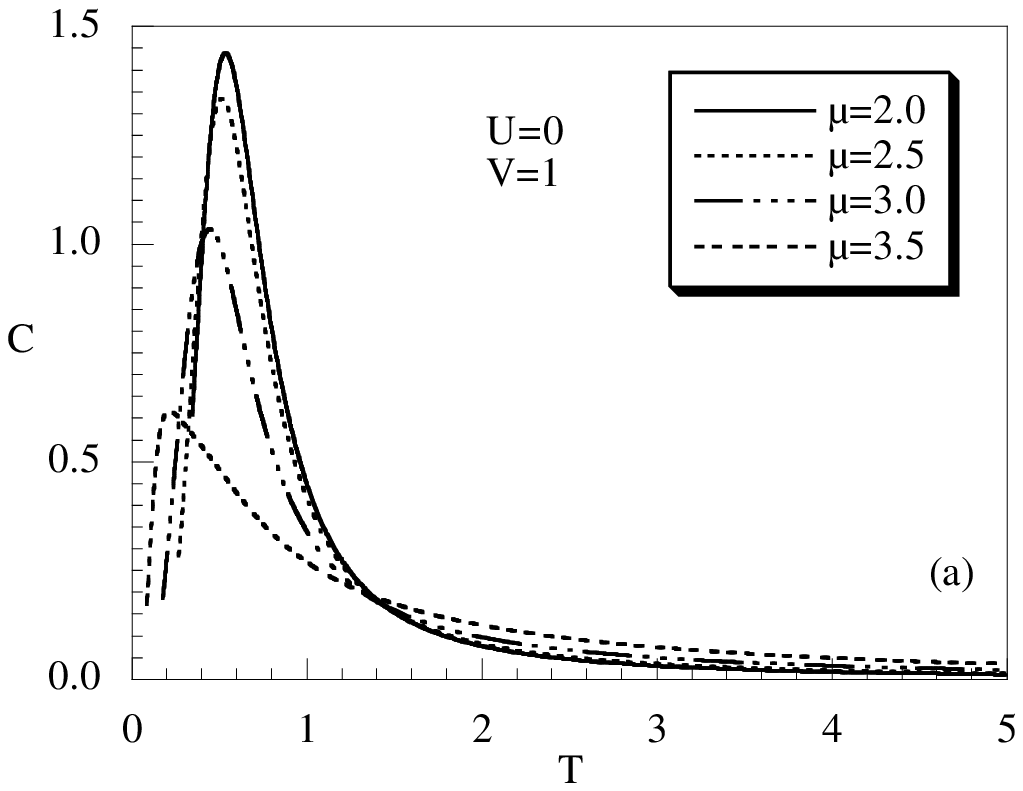}
\includegraphics[width=5.3cm]{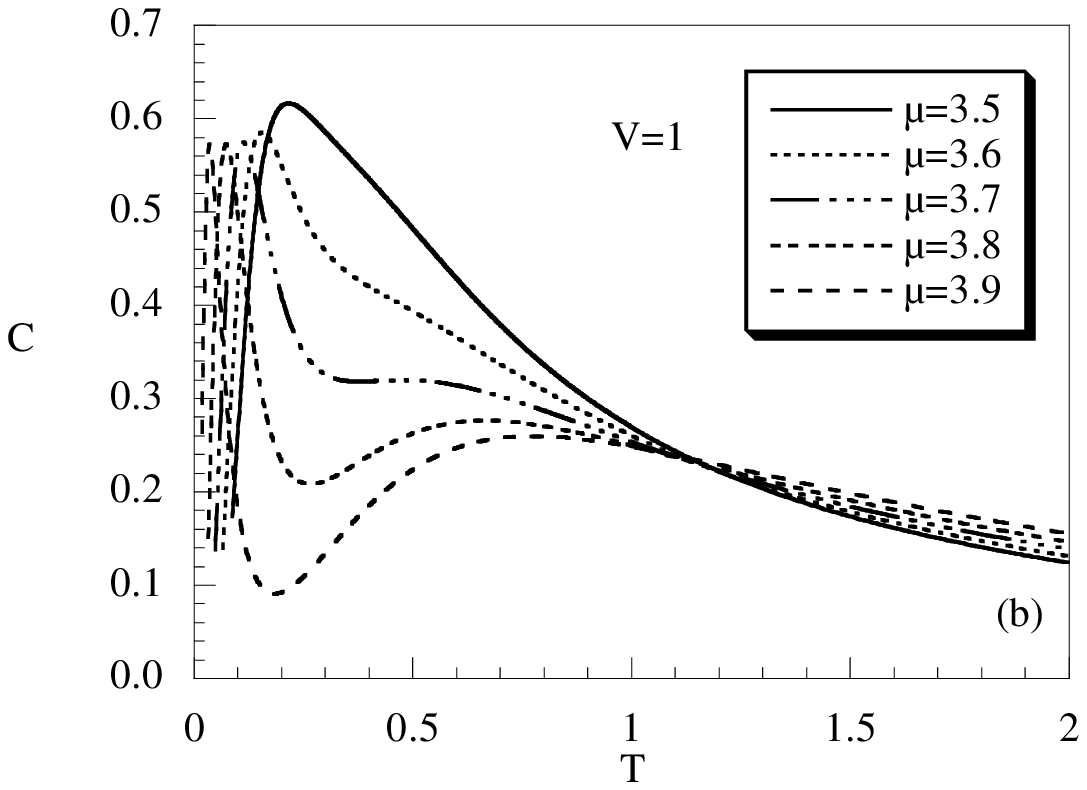}
\includegraphics[width=5.3cm]{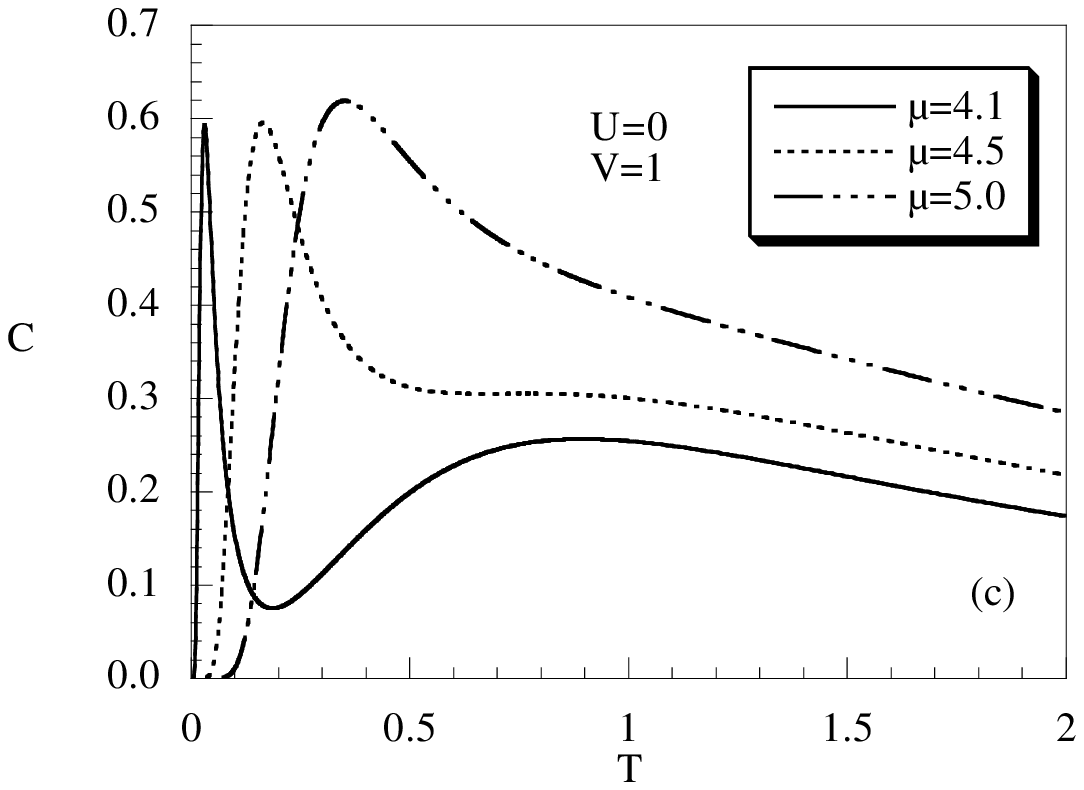}
\caption{The specific heat $C$ is plotted as a function of the
temperature at various values of chemical potential for $V=1$: (a)
in the region $2V<\mu<4V$; (b) near the transition point $\mu=4V$;
(c) in the region $\mu>4V $.} \label{Fig8}
\end{figure}


To study the specific heat, let us distinguish the two regions
$2V<\mu <4V$ and $\mu >4V$. In the first region, see Fig.~\ref{Fig8}
(a), the specific heat increases with temperature, exhibits a peak
at a certain temperature $T=T_{1}$, then decreases. When $\mu $
approaches the critical value $\mu _{c}=4V$, see Fig.~\ref{Fig8}
(b), the specific heat develops a double peak structure with a broad
peak at higher temperature than $T_{1}$. The latter temperature
decreases with $\mu $ and tends to zero for $\mu \rightarrow \mu
_{c}$. It is characteristic of this region the fact that all the
specific curves cross at the same temperature, independently on the
value of $\mu $. The crossing temperature is $T^{*}\approx 1.1$. The
fact that there is a crossing point in the specific heat curves
versus $T$, when plotted for different values of some thermodynamics
quantities, has been observed in a large variety of systems
\cite{Vollhardt97,Mancini99,Mancini04}. In the second region, see
Fig.~\ref{Fig8} (c), C exhibits a low-temperature peak at a certain
temperature $T=T_{2}$. $T_{2}$ tends to zero for $\mu \rightarrow
\mu _{c}$, increases by increasing $\mu $. Again, close to $\mu
_{c}$ there is a double-peak structure, which disappears when $\mu $
moves away from $\mu _{c}$. It is characteristic of this region that
no crossing point is observed. For large value of $T$, the two
specific heats (i.e. attractive and repulsive $V$) tend to coincide.
This is because the system is in a homogeneous phase, where the
thermal energy predominates over the Coulomb interaction. It is
interesting to observe that the crossing point is observed only for
$V>0$ and in the region $0<\mu <4V$, where the checkerboard order is
observed.

\begin{figure}[tb]
\includegraphics[width=8cm]{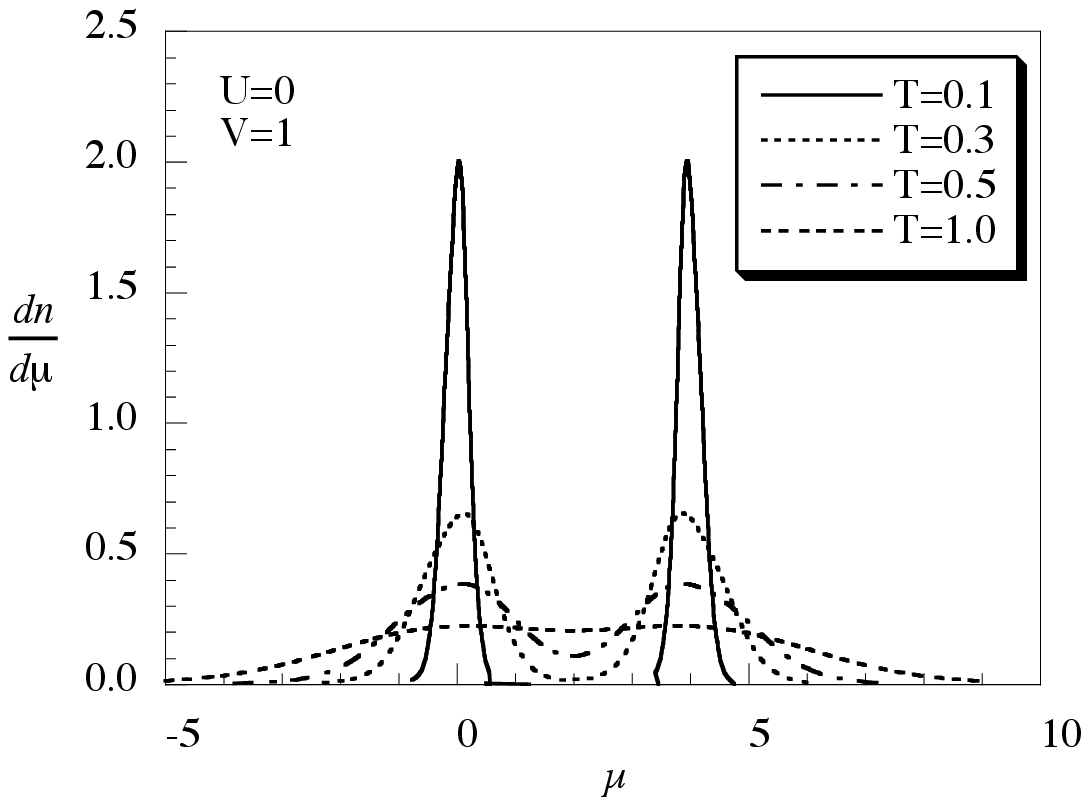}
\caption{The derivative of the particle density with respect to the
chemical potential $dn/d\mu$ is plotted as a function of the
chemical potential at various temperatures for $V=1$.} \label{Fig9}
\end{figure}


The thermal compressibility $\kappa ^{T}$ is studied in
Fig.~\ref{Fig9}, where $\frac{dn}{d\mu }$ is given as a function of
the chemical potential for various values of temperature. When $T$
is lowered a double peak structure develops, with peaks localized at
$\mu =0$ and $\mu =4V$. The heights of the
peaks increases by decreasing $T$ and tends to diverge in the limit $%
T\rightarrow 0$. It is worth noticing that for low temperature the
compressibility is very small (zero in the limit $T\rightarrow 0$)
in the wide region $0<\mu <4V$, where the phase with checkerboard
order of the charge is observed.\ Similar results have been obtained
in Ref.~\onlinecite{Zhang04}, where the $t-V$ model has been studied
within a cluster approximation.

\section{Conclusions}

The Hubbard model with intersite Coulomb interaction has been
studied in the ionic limit (i.e., no kinetic energy). This model is
isomorphic to the spin-1 Ising model in presence of a crystal field
and an external magnetic field. A finite complete set of
eigenoperators and eigenvalues of the Hamiltonian has been found for
arbitrary dimensions. This knowledge allows us to determine
analytical expressions of the local Green's functions and the
correlation functions. As the eigenoperators do not satisfy a
canonical algebra, the GF and the CF depend on a set of unknown
parameters, not calculable by means of the dynamics. By using
appropriate boundary conditions and algebraic relations we have
determined these parameters for the case of an infinite homogeneous
chain. Some results for the case $U=0$ (i.e., no local Coulomb
interaction / no crystal field) have been given. The system exhibits
a different behavior according to the sign of $V$, the intersite
Coulomb interaction, or the sign of $J$, the exchange interaction.
For $V<0$ ($J>0$, ferromagnetic coupling) at $\mu =2V$ ($h=0$) the
system exhibits a phase transition to a charge ordered state
(ferromagnetic phase for the Ising model) at zero temperature. For
positive $V$ ($J<0$, antiferromagnetic coupling), the system
exhibits instabilities at $\mu =0$ ($h=-2\left| J\right| $) and at
$\mu =4V$ ($h=2\left| J\right| $). In the limit of zero temperature
a phase transition to a state with a checkerboard order of the
charge (antiferromagnetic phase for the Ising model) is observed at
$\mu =0$.\ This order persists up to $\mu =4V$, where a second
transition to an homogeneous charge order is observed.\ A crossing
point in the specific heat curves is observed only for $V>0$ and in
the region $0<\mu <4V$, where the checkerboard order is observed. In
the entire region $0<\mu <4V$, the compressibility vanishes at low
temperatures. Further study, in particular for the case of finite
$U$, will be presented elsewhere.

\appendix

\section{Algebraic relations}

Let us start by observing that because of the basic anticommutating
rules (\ref{2.3}) the number $n(i)=c^{\dag }(i)c(i)$ and the double
occupancy $ D(i)=n(i)[n(i)-1]/2$ operators satisfy $for\;p\geq 1$
the following algebra
\begin{equation}
\begin{array}{l}
n^{p}(i)=n(i)+a_{p}D(i) \\
D^{p}(i)=D(i) \\
n^{p}(i)D(i)=2^{p}D(i)
\end{array}
\;\;\;\;\;\;a_{p}=2^{p}-2  \label{A1}
\end{equation}
From this algebra several and important relations can be derived.
Firstly, let us consider the operator
\begin{equation}
n^{\alpha
}(i)=\frac{1}{2d}\stackrel{2d}{\sum_{m=1}}n(\mathbf{i}_{m})
\label{A2}
\end{equation}
where $\mathbf{i}_{m}$ are the first neighbors of the site
$\mathbf{i}$. A
basic relation can be derived for the operator $[n^{\alpha }(i)]^{p}$, with $%
p\ge 1$. In this and in the following Appendices we shall present
results for the one-dimensional system \footnote{In Appendices A, B
and C, the calculations for the recursion rule, the energy and
spectral density matrices are reported for the one-dimensional case.
Calculations for higher dimensions can be made by using the same
technique. For interested readers, the results for $d=2$ and $3$ are
given in a technical report \cite{ManciniTC}, available on
request.}. We start from the equation
\begin{equation}
\lbrack n^{\alpha
}(i)]^{p}={\frac{1}{{2^{p}}}}\sum\limits_{m=0}^{p}{}\left(
\begin{array}{c}
p \\
m
\end{array}
\right) n(i_{1})^{p-m}n(i_{2})^{m}  \label{A3}
\end{equation}
Because of the algebraic relations (\ref{A1}) we obtain
\begin{equation}
\lbrack n^{\alpha }(i)]^{p}={\frac{1}{{2^{p}}}}\sum%
\limits_{m=1}^{4}{}b_{m}^{(p)}Z_{m}  \label{A4}
\end{equation}
where the operators $Z_{m}$ are defined as
\begin{equation}
\begin{array}{l}
{Z_{1}=2n^{\alpha }(i)} \\
{Z_{2}=2D^{\alpha }(i)+n(i_{1})n(i_{2})} \\
{Z_{3}=D(i_{1})n(i_{2})+D(i_{2})n(i_{1})} \\
{Z_{4}=D(i_{1})D(i_{2})}
\end{array}
\label{A5}
\end{equation}
and the coefficients $b_{m}^{(p)}$ have the expressions
\begin{equation}
\begin{array}{l}
{b_{1}^{(p)}=1} \\
{b_{2}^{(p)}=\sum\limits_{m=1}^{p-1}{}\left(
\begin{array}{c}
p \\
m
\end{array}
\right) =2(-1+2^{p-1})} \\
{b_{3}^{(p)}=\sum\limits_{m=1}^{p-1}{}\left(
\begin{array}{c}
p \\
m
\end{array}
\right) a_{p-m}=3(1-2^{p}+3^{p-1})} \\
{b_{4}^{(p)}=\sum\limits_{m=1}^{p-1}{}\left(
\begin{array}{c}
p \\
m
\end{array}
\right) a_{p-m}a_{m}=4(-1+3\cdot 2^{p-1}-3^{p}+4^{p-1})}
\end{array}
\label{A6}
\end{equation}
By solving the system (\ref{A4}) with respect to the variables
$Z_{m}$, we can obtain from (\ref{A4}) the recurrence rule
\begin{equation}
\lbrack n^{\alpha
}(i)]^{p}=\sum\limits_{m=1}^{4}{}A_{m}^{(p)}[n^{\alpha }(i)]^{m}
\label{A7}
\end{equation}
where the coefficients $A_{m}^{(p)}$ are defined as
\begin{equation}
\begin{array}{l}
{A_{1}^{(p)}={\frac{1}{{2^{p-1}}}}[b_{1}^{(p)}-{\frac{1}{2}}b_{2}^{(p)}+{\
\frac{1}{3}}b_{3}^{(p)}-{\frac{1}{4}}b_{4}^{(p)}]} \\
{A_{2}^{(p)}={\frac{1}{{2^{p-1}}}}[b_{2}^{(p)}-b_{3}^{(p)}+{\frac{{11}}{{12}}%
}b_{4}^{(p)}]} \\
{A_{3}^{(p)}={\frac{1}{{2^{p-1}}}}[{\frac{2}{3}}b_{3}^{(p)}-b_{4}^{(p)}]} \\
{A_{4}^{(p)}={\frac{1}{{2^{p-1}}}}{\frac{1}{3}}b_{4}^{(p)}}
\end{array}
\label{A8}
\end{equation}
We note that
\begin{equation}
\sum\limits_{m=1}^{4}{}A_{m}^{(p)}=1  \label{A9}
\end{equation}
In Table 1 we give some values of the coefficients $A_{m}^{(p)}.$

$
\begin{tabular}{|c|c|c|c|c|}
\hline $p$ & $A_{1}^{(p)}$ & $A_{2}^{(p)}$ & $A_{3}^{(p)}$ &
$A_{4}^{(p)}$ \\ \hline $1$ & $1$ & $0$ & $0$ & $0$ \\ \hline $2$ &
$0$ & $1$ & $0$ & $0$ \\ \hline $3$ & $0$ & $0$ & $1$ & $0$ \\
\hline $4$ & $0$ & $0$ & $0$ & $1$ \\ \hline $5$ & $-\frac{3}{2}$ &
$\frac{25}{4}$ & $-\frac{35}{4}$ & $5$ \\ \hline $6$ &
$-\frac{15}{2}$ & $\frac{119}{4}$ & $-\frac{75}{2}$ & $\frac{65}{4}$
\\ \hline
$7$ & $-\frac{195}{8}$ & $\frac{1505}{16}$ & $-\frac{1799}{16}$ &
$\frac{175 }{4}$ \\ \hline $8$ & $-\frac{525}{8}$ &
$\frac{3985}{16}$ & $-\frac{1155}{4}$ & $\frac{1701 }{16}$ \\ \hline
\end{tabular}
$

\section{The energies matrices}

The energy matrices $\epsilon ^{(\xi )}$ and $\epsilon ^{(\eta )}$
can be calculated by means of the equations of motion (\ref{3.4})
and (\ref{3.5}) and the recurrence rule (\ref{3.7}). The explicit
expressions are
\begin{equation}
\varepsilon ^{(\xi )}=\left(
\begin{array}{ccccc}
{-\mu } & {2dV} & {\cdots } & 0 & 0 \\
0 & {-\mu } & {\cdots } & 0 & 0 \\
\vdots & \vdots & \vdots & \vdots & \vdots \\
0 & 0 & {\cdots } & {-\mu } & {2dV} \\
0 & {{{2dVA_{1}^{(4d+1)}}}} & {\cdots } & {{{2dVA_{4d-1}^{(4d+1)}}}} & {{{%
-\mu +2dVA_{4d}^{(4d+1)}}}}
\end{array}
\right)  \label{B01}
\end{equation}
\begin{equation}
\varepsilon ^{(\eta )}=\left(
\begin{array}{ccccc}
U{-\mu } & {2dV} & {\cdots } & 0 & 0 \\
0 & U{-\mu } & {\cdots } & 0 & 0 \\
\vdots & \vdots & \vdots & \vdots & \vdots \\
0 & 0 & {\cdots } & U{-\mu } & {2dV} \\
0 & {{{2dVA_{1}^{(4d+1)}}}} & {\cdots } & {{{2dVA_{4d-1}^{(4d+1)}}}}
& U{{{\ -\mu +2dVA_{4d}^{(4d+1)}}}}
\end{array}
\right)  \label{B.02}
\end{equation}
where ${{{A_{m}^{(4d+1)}}}}$ are the coefficients appearing in
(\ref{3.4}). In particular, for one dimension

\begin{equation}
\epsilon ^{(\xi )}=\left(
\begin{array}{ccccc}
-\mu & 2V & 0 & 0 & 0 \\
0 & -\mu & 2V & 0 & 0 \\
0 & 0 & -\mu & 2V & 0 \\
0 & 0 & 0 & -\mu & 2V \\
0 & -3V & \frac{25V}{2} & -\frac{35}{2}V & -\mu +10V
\end{array}
\right)  \label{B1}
\end{equation}

\begin{equation}
\epsilon ^{(\eta )}=\left(
\begin{array}{ccccc}
U-\mu & 2V & 0 & 0 & 0 \\
0 & U-\mu & 2V & 0 & 0 \\
0 & 0 & U-\mu & 2V & 0 \\
0 & 0 & 0 & U-\mu & 2V \\
0 & -3V & \frac{25V}{2} & -\frac{35}{2}V & U-\mu +10V
\end{array}
\right)  \label{B2}
\end{equation}

The matrices $\Omega ^{(\xi )}$ and $\Omega ^{(\eta )}$ have the
expressions
\begin{equation}
\Omega ^{(\xi )}=\Omega ^{(\eta )}=\left(
\begin{array}{ccccc}
1 & 2^{4} & 1 & (2/3)^{4} & (1/2)^{4} \\
0 & 2^{3} & 1 & (2/3)^{3} & (1/2)^{3} \\
0 & 2^{2} & 1 & (2/3)^{2} & (1/2)^{2} \\
0 & 2 & 1 & (2/3) & (1/2) \\
0 & 1 & 1 & 1 & 1
\end{array}
\right)  \label{B3}
\end{equation}

\section{The spectral density matrices}

By means of the formulas (\ref{4.12}) and recalling the expressions
[see Appendix B] for the energy matrices, and the expressions
(\ref{4.6})-(\ref{4.10}) of the normalization matrix, we can easily
calculate the spectral
matrices. Furthermore, we note that the matrices $\Omega ^{(\xi )}$ and $%
\Omega ^{(\eta )}$ are equal. Then, the spectral density matrices
have similar form in terms of the matrix elements of the
normalization matrices. Because of the recurrence relation, we need
to calculate only the first row of the matrices. Calculations show
that the matrices $\sigma ^{(a,n)}$ have the following form
\begin{equation}
\sigma ^{(a,n)}=\Sigma _{n}^{(a)}\Gamma ^{(n)}\qquad n=1,2.\cdots
,4d+1 \label{D1}
\end{equation}
where $\Sigma _{n}^{(a)}$ are functions of the elements $I_{1,m}^{(a)}$ $%
(m=1,2.\cdots ,4d+1)$ and $\Gamma ^{(n)}$ are matrices of rank
$(4d+1)\times (4d+1)$. For the one-dimensional case

\begin{equation}
\begin{array}{l}
{\Sigma _{1}^{(a)}={\frac{1}{6}}%
(6I_{1,1}^{(a)}-25I_{1,2}^{(a)}+35I_{1,3}^{(a)}-20I_{1,4}^{(a)}+4I_{1,5}^{(a)})%
} \\
{\Sigma _{2}^{(a)}={\frac{4}{3}}%
(6I_{1,2}^{(a)}-13I_{1,3}^{(a)}+9I_{1,4}^{(a)}-2I_{1,5}^{(a)})} \\
{\Sigma
_{3}^{(a)}=-6I_{1,2}^{(a)}+19I_{1,3}^{(a)}-16I_{1,4}^{(a)}+4I_{1,5}^{(a)}}
\\
{\Sigma _{4}^{(a)}={\frac{4}{3}}%
(2I_{1,2}^{(a)}-7I_{1,3}^{(a)}+7I_{1,4}^{(a)}-2I_{1,5}^{(a)})} \\
{\Sigma _{5}^{(a)}={\frac{1}{6}}%
(-3I_{1,2}^{(a)}+11I_{1,3}^{(a)}-12I_{1,4}^{(a)}+4I_{1,5}^{(a)})}
\end{array}
\label{D2}
\end{equation}
\begin{equation}
\begin{array}{l}
\Gamma _{1,m}^{(1)}=(
\begin{array}{lllll}
1 & 0 & 0 & 0 & 0
\end{array}
) \\
\Gamma _{1,m}^{(2)}=(
\begin{array}{lllll}
1 & 2^{-1} & 2^{-2} & 2^{-3} & 2^{-4}
\end{array}
) \\
\Gamma _{1,m}^{(3)}=(
\begin{array}{lllll}
1 & 1 & 1 & 1 & 1
\end{array}
) \\
\Gamma _{1,m}^{(4)}=(
\begin{array}{lllll}
1 & (2/3)^{-1} & (2/3)^{-2} & (2/3)^{-3} & (2/3)^{-4}
\end{array}
) \\
\Gamma _{1,m}^{(5)}=(
\begin{array}{lllll}
1 & (1/2)^{-1} & (1/2)^{-2} & (1/2)^{-3} & (1/2)^{-4}
\end{array}
)
\end{array}
\label{D3}
\end{equation}


\end{document}